\documentclass[a4paper, 14pt, conference,onecolumn]{ieeeconf}      

\IEEEoverridecommandlockouts                              

\usepackage{graphics} 
\usepackage{epsfig} 

\usepackage{amsmath} 
\usepackage{amssymb}  
\usepackage{tabularx}
\usepackage{amsmath}
\usepackage{balance}

\newcounter{subassumption}[asu]

\makeatletter
\renewcommand{\p@subassumption}{\theasu}
\makeatother

\usepackage{rotating,booktabs,multirow}
\makeatletter
\let\NAT@parse\undefined
\makeatother
\usepackage{hyperref}
\usepackage{graphicx}
\newtheorem{remark}{Remark}[section]

\usepackage[latin1]{inputenc}
\usepackage{color}
\usepackage{epsfig}
\usepackage{graphicx}
\usepackage[hang]{caption2}
\usepackage[normalsize,it,hang]{subfigure}
\DeclareGraphicsExtensions{.jpg,.png,.jpg,.jpg}

\title{\LARGE \bf
Model-Free Control as a Service \\ in the Industrial Internet of Things: \\ Packet loss and latency issues \\ via preliminary experiments 
}

\author{C\'{e}dric Join$^{1, 3}$,  Michel Fliess$^{2, 3}$, Fr\'{e}d\'{e}ric Chaxel$^{1}$ 
\thanks{$^1$CRAN (CNRS, UMR 7039)), Universit\'{e} de Lorraine, BP 239, 54506 Vand{\oe}uvre-l\`{e}s-Nancy, France. \newline
{\tt\small \{Cedric.Join, Frederic.Chaxel\}@\newline univ-lorraine.fr}}
\thanks{$^{2}$LIX (CNRS, UMR 7161), \'Ecole polytechnique, 91128 Palaiseau, France. {\tt \small Michel.Fliess@polytechnique.edu } } 
\thanks{$^{3}$AL.I.E.N. (ALg\`ebre pour Identification \& Estimation Num\'eriques), 7 rue Maurice Barr\`{e}s, 54330 V\'{e}zelise, France. \newline
        {\tt \small \{cedric.join, michel.fliess\}@alien-sas.com}}
        }

\begin{document}
\maketitle
\thispagestyle{empty}
\pagestyle{empty}

\begin{abstract}
Model-Free Control (MFC), which is easy to implement both from software and hardware viewpoints, permits the introduction of a high level control synthesis for the Industrial Internet of Things (IIoT) and the Industry 4.0. The choice of the User Diagram Protocol (UDP) as the Internet Protocol permits to neglect the latency. In spite of most severe packet losses, convincing computer simulations and laboratory experiments show that MFC exhibits a good Quality of Service (QoS) and behaves better than a classic PI regulator.

\keywords Control engineering, model-free control, intelligent controllers, industrial internet of things, industry 4.0, cyber-physical systems, cloud computing, latency, packet loss, congestion, UDP protocol, half quadrotor, joystick.
\end{abstract}

\section{Introduction}
The conceptual meanings of the buzzwords \emph{Industrial Internet of Things} (\emph{IIoT}) (see, \textit{e.g.}, \cite{boyes,kim}),  \emph{industry 4.0} (see, \textit{e.g.}, \cite{gilchrist,schulz}), \emph{cyber-physical systems} (see, \textit{e.g.}, \cite{manfredi}) do overlap to some extent (see, \textit{e.g}, \cite{gilchrist,hermann}). \emph{Control engineering} (see, \textit{e.g.}, \cite{murray}) plays there a key r\^{o}le (see, \textit{e.g.}, \cite{xia1,zhang}) via networks that are often related to \emph{cloud computing} (see, \textit{e.g.}, \cite{liu,marinescu}).  

Among the numerous existing control strategies, \emph{model predictive control} (see, \textit{e.g.}, \cite{camacho}) seems today the most popular one, at least in the academic literature (see, \textit{e.g.}, \cite{abdel,arzen,carli,ha,kouki1,kouki2,megahed,skarin,vick,xia2}). This communication advocates \emph{Model-Free Control} (\emph{MFC}) in the sense of \cite{csm}, and the corresponding ``intelligent'' controllers. This setting, which is easy to implement both from software \cite{csm} and hardware \cite{hardware} viewpoints, will hopefully lead in some near future to \emph{Model-Free Control as a Service} ({\em MFCaaS}). It has been already most successfully applied in many concrete situations (see the references in \cite{csm} and \cite{bara} for a quite complete listing until the beginning of 2018). Some have been patented. The recent contributions of MFC to the dynamic adaptation of computing resource allocations under time-varying workload in cloud computing \cite{bekcheva} and to the air-conditioning of data centers \cite{hvac} should be emphasized here. 

The choice of an appropriate \emph{Internet Protocol} (\emph{IP}) stack is of utmost importance in this networking context (see, \textit{e.g.}, \cite{kurose}). It is obvious that packet loss and latency, which are unavoidable, might significantly degrade the performances of any control law.  There are two main protocols of transport layer, the \emph{Transmission  Control Procol} (\emph{TCP}) and the \emph{User Datagram Procol} (\emph{UDP}) (see, \textit{e.g.}, \cite{kumar,udp2,udp1} for some details). TCP is more reliable but may exhibit often fatal latency and jitter. This is why we select here UDP, which is faster: 
\begin{itemize}
\item It permits to neglect the delay if the transmission distance is not ``too'' large (see Section \ref{conclusion}). 
\item Only packet loss, which might be most severe, is taken into account (compare, \textit{e.g.}, with \cite{millnert}). 
\item Packets that arrive late are discarded.
\item Congestion may therefore be somehow ignored.
\end{itemize}
This communication, which completes a recent technical report \cite{hal}, is organized as follows. Basic facts about MFC are summarized in Section \ref{free}. Section \ref{exper} is devoted to computer simulations. After the introduction of two types of packet loss in Section \ref{gen},  a single tank is analyzed in Section \ref{tank}: the computer simulations for MFC indicate in spite of serious packet losses a fine \emph{Quality of Service} (\emph{QoS}), which is  much better than with a classic PI. Those ascertainments are confirmed in Section \ref{quanser} via laboratory experiments  with the Quanser AERO, \textit{i.e.}, a half quadrotor. In Section \ref{joystick} a joystick is added. See Section \ref{conclusion} for some suggestions on prospective studies.

\section{Model-free control and intelligent controllers\protect\footnote{See \cite{csm} for more details.}} \label{free}
\subsection{The ultra-local model and intelligent controllers}
For the sake of notational simplicity, let us restrict ourselves to single-input single-output (SISO) systems. The unknown global description of the plant is replaced by the following first-order \emph{ultra-local model}:
\begin{equation}
\boxed{\dot{y} = F + \alpha u} \label{1}
\end{equation}
where
\begin{enumerate}
\item the control and output variables are respectively $u$ and $y$,
\item $\alpha \in \mathbb{R}$ is chosen by the practitioner such that the three terms in Equation \eqref{1} $\alpha u$ are of the same magnitude.
\end{enumerate}
The following comments are useful:
\begin{itemize}
\item $F$ is \emph{data driven}: it is given by the past values of $u$ and $y$.
\item $F$ includes not only the unknown structure of the system but also any disturbance.
\end{itemize}




Close the loop with the \emph{intelligent proportional controller}, or \emph{iP}, 
\begin{equation}\label{ip}
\boxed{u = - \frac{F_{\text{est}} - \dot{y}^\ast + K_P e}{\alpha}}
\end{equation}
where
\begin{itemize}
\item $y^\ast$ is the reference trajectory,
\item $e = y - y^\star$ is the tracking error,
\item $F_{\text{est}}$ is an estimated value of $F$,
\item $K_P \in \mathbb{R}$ is a gain.
\end{itemize}
Equations \eqref{1} and \eqref{ip} yield
\begin{equation}
	\dot{e} + K_P e = F - F_{\text{est}}
			\label{equa15}
	\end{equation}
If the estimation $F_{\text{est}}$ is ``good'': $F - F_{\text{est}}$ is ``small'', \textit{i.e.}, $F - F_{\text{est}} \simeq 0$,  then $\lim_{t \to +\infty} e(t) \simeq 0$ if $K_P > 0$. It implies that the tuning of $K_P$ is quite straightforward. This is a major benefit when
compared to the tuning of ``classic'' PIDs (see, \textit{e.g.},
\cite{murray,broida}). 
\begin{remark}
See \cite{csm,join} for other types of ultra-local models, where the derivation order of $y$ in Equation \eqref{1} should be greater than $1$, and for the corresponding intelligent controllers. The extension to MIMO systems is straightforward \cite{toulon}.
\end{remark}

\subsection{Estimation of $F$}\label{F}
Mathematical analysis (see, \textit{e.g.}, \cite{bourbaki}) tells us that under a very weak integrability assumption, any function, for instance $F$ in Equation \eqref{1}, is ``well'' approximated by a piecewise constant function. The estimation techniques below are borrowed 
from \cite{sira1,sira}.
\subsubsection{First  approach}
Rewrite then Equation \eqref{1}  in the operational domain (see, \textit{e.g.}, \cite{yosida}): 

\begin{equation}
sY = \frac{\Phi}{s}+\alpha U +y(0)
			\label{equa16}
\end{equation}
where $\Phi$ is a constant. We get rid of the initial condition $y(0)$ by multiplying both sides on the left by $\frac{d}{ds}$:
\begin{equation}
Y + s\frac{dY}{ds}=-\frac{\Phi}{s^2}+\alpha \frac{dU}{ds}
			\label{equa17}
	\end{equation}
Noise attenuation is achieved by multiplying both sides on the left by $s^{-2}$. It yields in the time domain the real-time estimate, thanks to the equivalence between $\frac{d}{ds}$ and the multiplication by $-t$,
\begin{equation*}\label{integral}
{\small F_{\text{est}}(t)  =-\frac{6}{\tau^3}\int_{t-\tau}^t \left\lbrack (\tau -2\sigma)y(\sigma)+\alpha\sigma(\tau -\sigma)u(\sigma) \right\rbrack d\sigma }
\end{equation*}
where $\tau > 0$ might be quite small. This integral, which is a low pass filter, may of course be replaced in practice by a classic digital linear filter.

\subsubsection{Second approach}\label{2e}
Close the loop with the iP \eqref{ip}. It yields:

\begin{equation*}
F_{\text{est}}(t) = \frac{1}{\tau}\left[\int_{t - \tau}^{t}\left(\dot{y}^{\star}-\alpha u
- K_P e \right) d\sigma \right] 
			\label{equa18}
	\end{equation*}

\section{Computer experiments}\label{exper}
\subsection{Generalities}\label{gen}
We use an intelligent proportional controller, \textit{i.e.}, Formula \eqref{ip}, where $F$ and $u$ are obtained thanks to a computer server which is connected to the plant via UDP. 
Two types of packet loss are considered :
\begin{itemize}
\item {\bf Fault 1} -- Some measurements of the sensor $y$ do not reach the server. The estimation of $F$ and $u$ is frozen.
\item {\bf Fault 2} -- The calculations of the server do not reach the plant. The control variable $u$ is thus frozen, but not the estimation of $F$.
\end{itemize}

\subsection{A single tank}\label{tank}
\subsubsection{Model-free control}
The following mathematical model is only useful for computer simulations:\footnote{See the real-time Matlab example: \newline {\tt \scriptsize https://fr.mathworks.com/help/sldrt/ug/\\water-tank-model-with-dashboard.html?s\_tid=srchtitle}}
\begin{equation} \label{cuve}
\dot y = \frac{\left( u-0.2700 \frak{K} \sqrt{y}\right)}{5} \quad 0 < y < 60, 0 < u < 70
\end{equation}
The outlet valve opening $\frak{K} $, $0 < \frak{K}  < 100$, should be viewed as an unknown perturbation.  The output is corrupted by an additive band-limited white noise of power $0.025$ (see, \textit{e.g.},\cite{siebert}). The sampling time is $100$ms. The simulations duration is equal to $200$s. The reference trajectory $y^\ast$, which is piecewise constant, explores all the possibilities: $y^\ast (t) = 0$ if $0 \leq t < 10$s, $y^\ast (t) = 15$ if $t < 10 \leq t < 80$s, $y^\ast (t) = 40$ if $80 \leq t < 100$s,  $y^\ast (t) = 55$ if $100 \leq t < 130$s, $y^\ast (t) = 10$ if $130 \leq t < 180$s, $y^\ast (t) = 0$ if $180 \leq t < 200$s. Set  $\frak{K}  = 10$ if $0 \leq t < 30$,  $\frak{K}  = 50$ if $30 \leq t < 120$,  $\frak{K}  = 20$ if $120 \leq t < 200$. Set in Formula \eqref{ip} $\alpha = 0.1$, $K_P = 0.5$. In order to assess the effects of the packet loss 5 scenarios are considered:
\begin{itemize}
\item {\bf Scenario 1} -- Tracking of the reference trajectory and no packet loss.
\item {\bf Scenario 2} -- Fault 1 (resp. 2) occurs if $140 \leq t < 150$ (resp. $50 \leq t < 60$).
\item {\bf Scenarios 3, 4 \& 5} -- There is $30\%$ (resp. $50\%$, $70\%$) of packet loss. Both types are evenly distributed
\end{itemize}

\begin{figure*}[!ht]
\centering%
\subfigure[\footnotesize PI: output variable (red), setpoint (black) and reference trajectory (blue)]
{\epsfig{figure=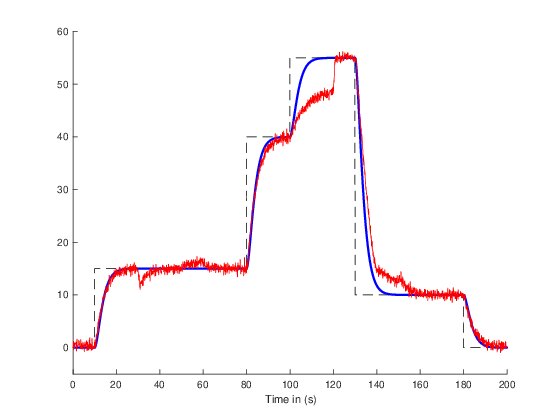,width=0.2700\textwidth}}
\subfigure[\footnotesize Control variable ]
{\epsfig{figure=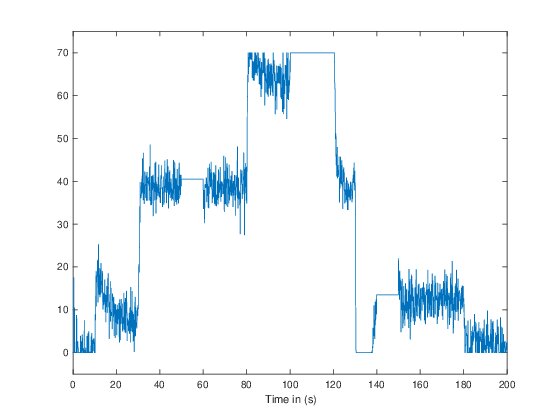,width=0.2700\textwidth}}
\subfigure[\footnotesize 0: no loss, 1: fault 1, 2: fault 2 ]
{\epsfig{figure=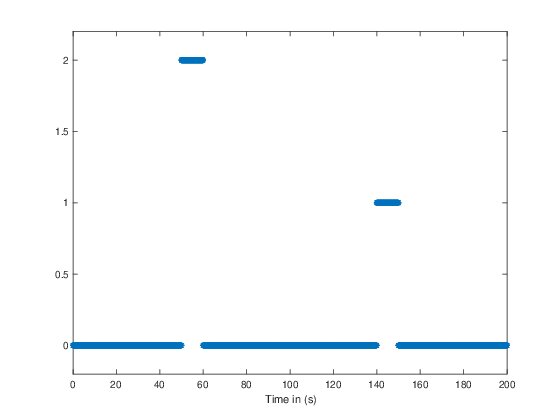,width=0.2700\textwidth}}
\caption{Scenario 2: MFC}\label{Y2}
\end{figure*}

\begin{figure*}[!ht]
\centering%
\subfigure[\footnotesize PI: output variable (red), setpoint (black) and reference trajectory (blue)]
{\epsfig{figure=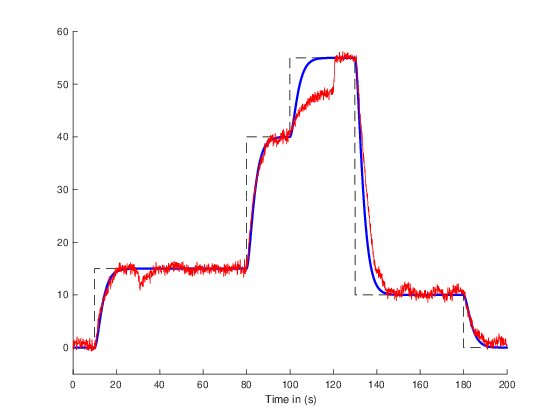,width=.2700\textwidth}}
\subfigure[\footnotesize Control variable]
{\epsfig{figure=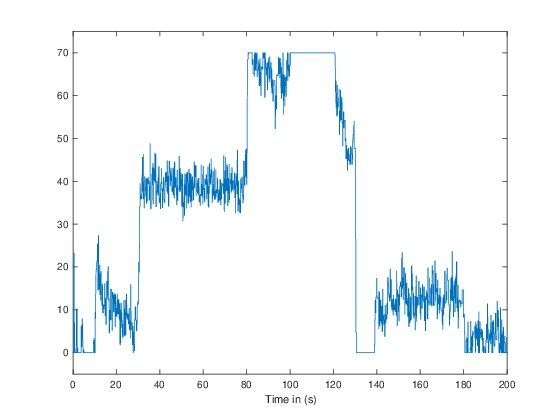,width=.2700\textwidth}}
\subfigure[\footnotesize Zoom on the faults]
{\epsfig{figure=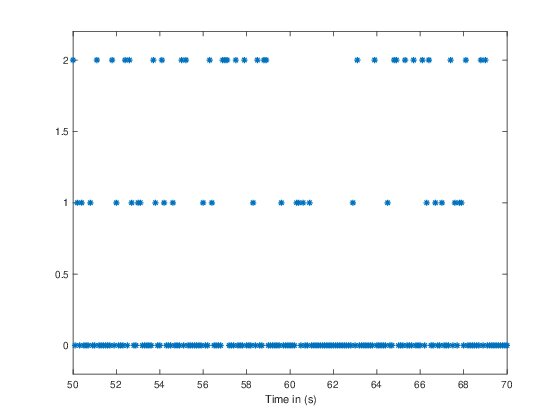,width=.2700\textwidth}}\\
\centering%
\subfigure[\footnotesize PI: output variable (red), setpoint (black) and reference trajectory (blue)]
{\epsfig{figure=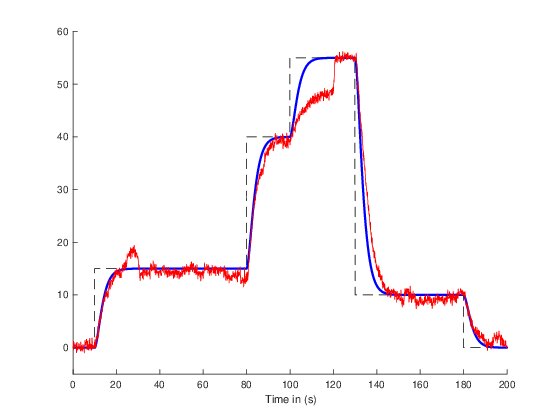,width=.2700\textwidth}}
\subfigure[\footnotesize Control variable]
{\epsfig{figure=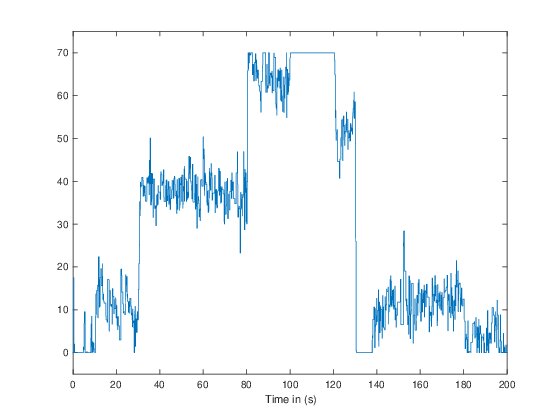,width=.2700\textwidth}}
\subfigure[\footnotesize Zoom on the faults]
{\epsfig{figure=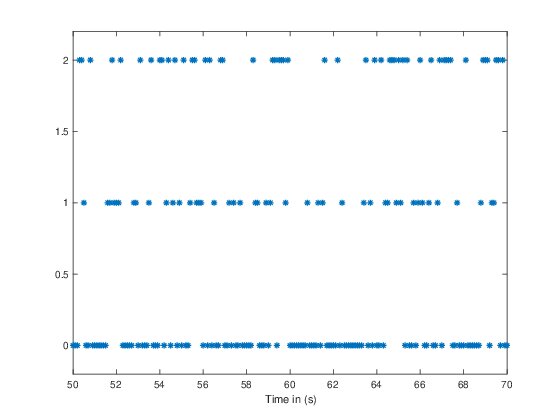,width=.2700\textwidth}}\\
\centering%
\subfigure[\footnotesize PI: output variable (red), setpoint (black) and reference trajectory (blue)]
{\epsfig{figure=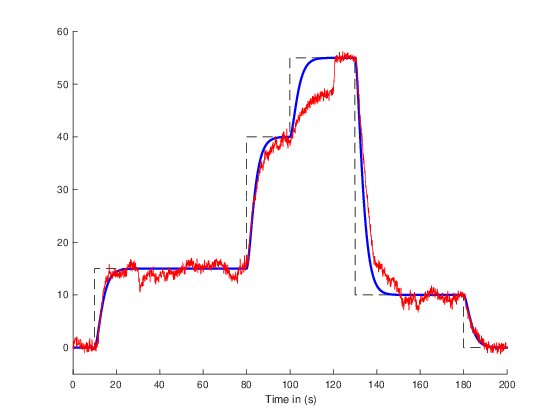,width=.2700\textwidth}}
\subfigure[\footnotesize Control variable]
{\epsfig{figure=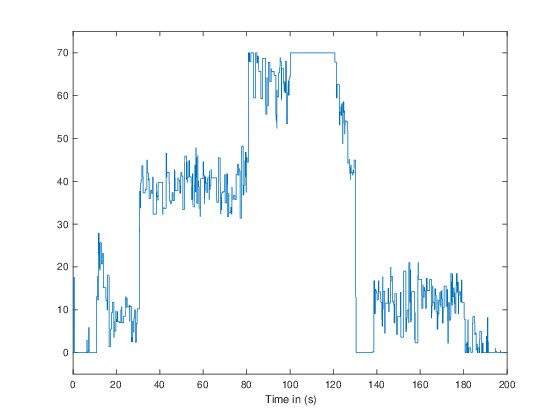,width=.2700\textwidth}}
\subfigure[\footnotesize Zoom on the faults]
{\epsfig{figure=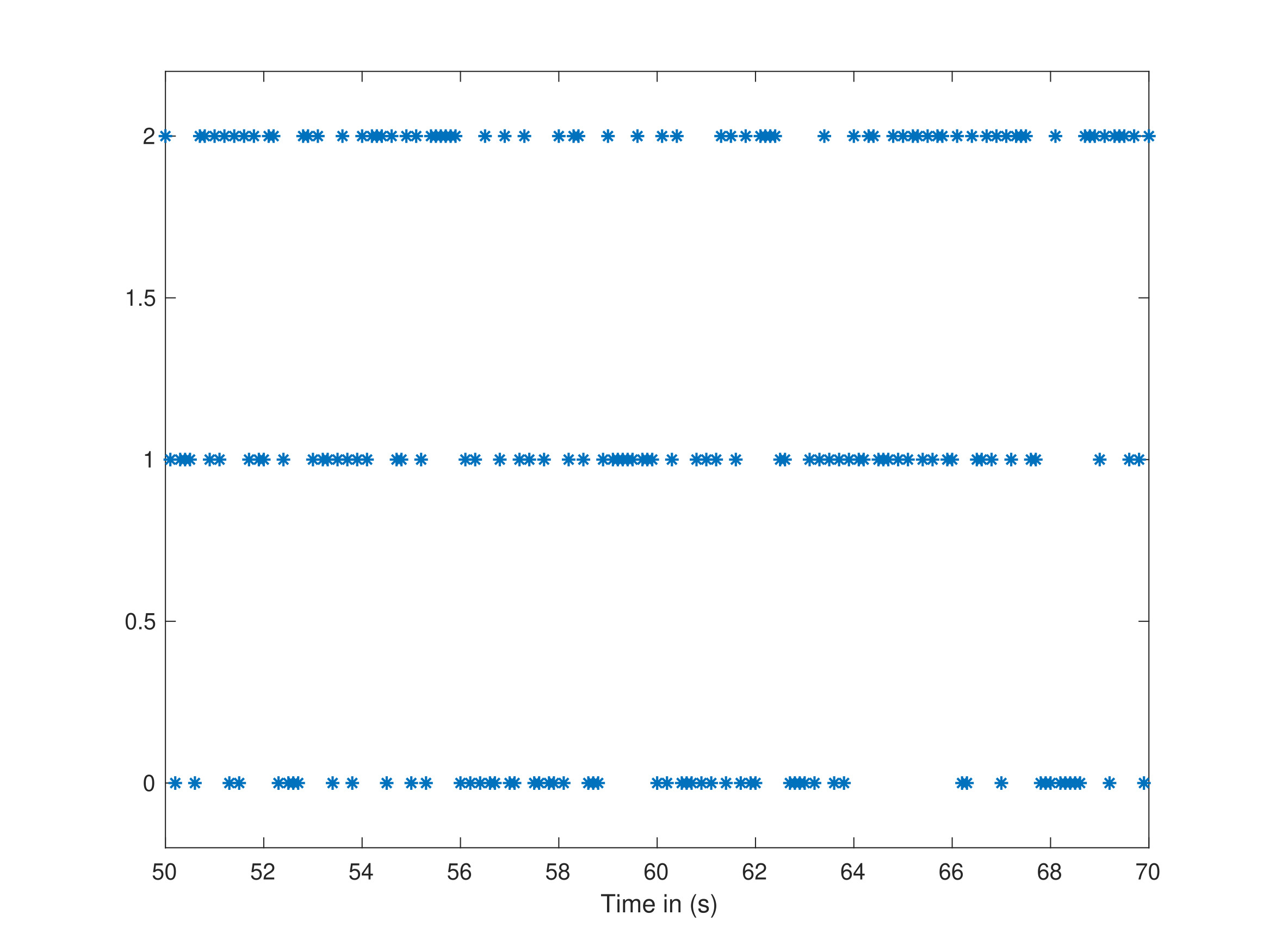,width=.2700\textwidth}}
\caption{Scenarios 3, 4 \& 5: MFC}\label{Y3}
\end{figure*}

\noindent Figures \ref{Y1}-\ref{Y3} display strong performances in spite of a big packet loss and significant variations of the parameter $\frak{K}$. The poor tracking of the setpoint when  $100 < t < 120$ is due to the saturation of control variable $u$ and not to the packet loss.
\subsubsection{A comparison with a PI controller}
\begin{figure*}[!ht]
\centering%
\subfigure[\footnotesize MFC: output variable (red), setpoint (black) and reference trajectory (blue)]
{\epsfig{figure=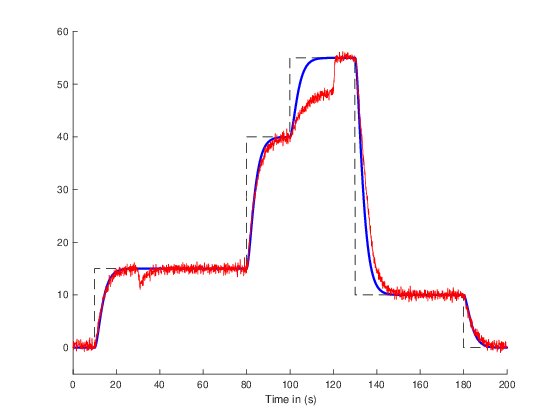,width=.2700\textwidth}}
\subfigure[\footnotesize MFC: control variable ]
{\epsfig{figure=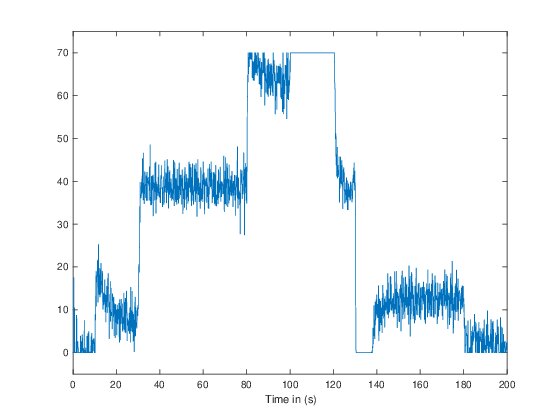,width=.2700\textwidth}}
\centering%
\subfigure[\footnotesize PI: output variable (red), setpoint (black) and reference trajectory (blue)]
{\epsfig{figure=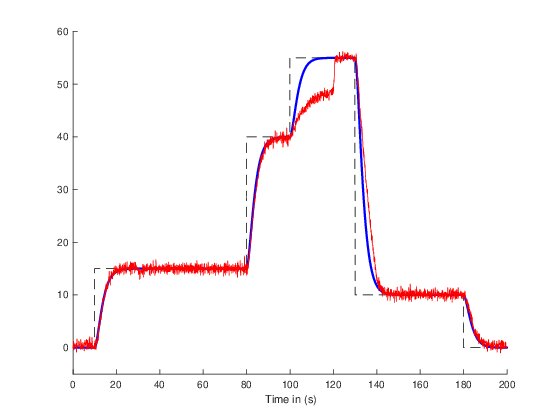,width=.2700\textwidth}}
\subfigure[\footnotesize PI: control variable]
{\epsfig{figure=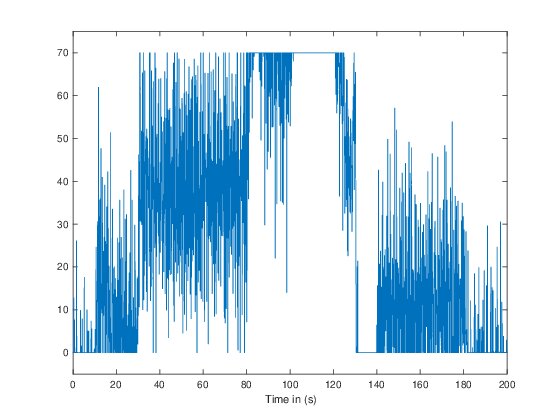,width=.2700\textwidth}}
\caption{Scénario 1: MFC \& PI}\label{Y1}
\end{figure*}

\begin{figure*}[!ht]
\centering%
\subfigure[\footnotesize PI: output variable (red), setpoint (black) and reference trajectory (blue)]
{\epsfig{figure=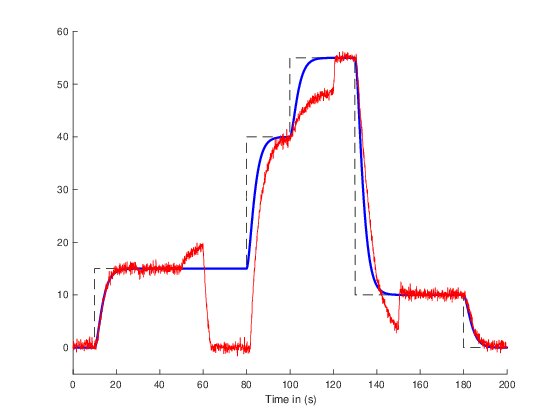,width=.2700\textwidth}}
\subfigure[\footnotesize Control variable]
{\epsfig{figure=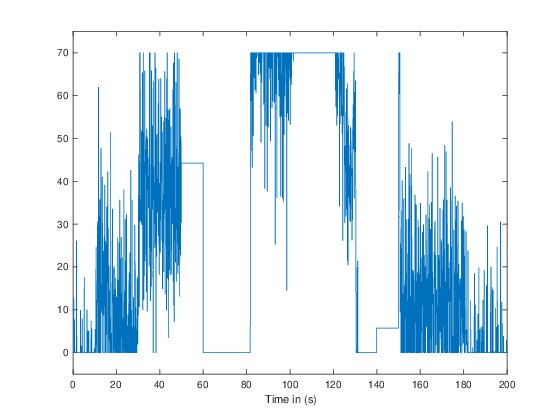,width=.2700\textwidth}}
\subfigure[\footnotesize  0: no loss, 1: fault 1, 2: fault 2 ]
{\epsfig{figure=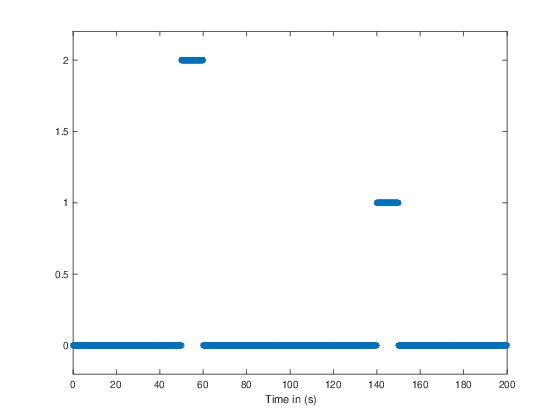,width=.2700\textwidth}}
\caption{Scenario 2: PI}\label{Y2P}
\end{figure*}

\begin{figure*}[!ht]
\centering%
\subfigure[\footnotesize PI: output variable (red), setpoint (black) and reference trajectory (blue)]
{\epsfig{figure=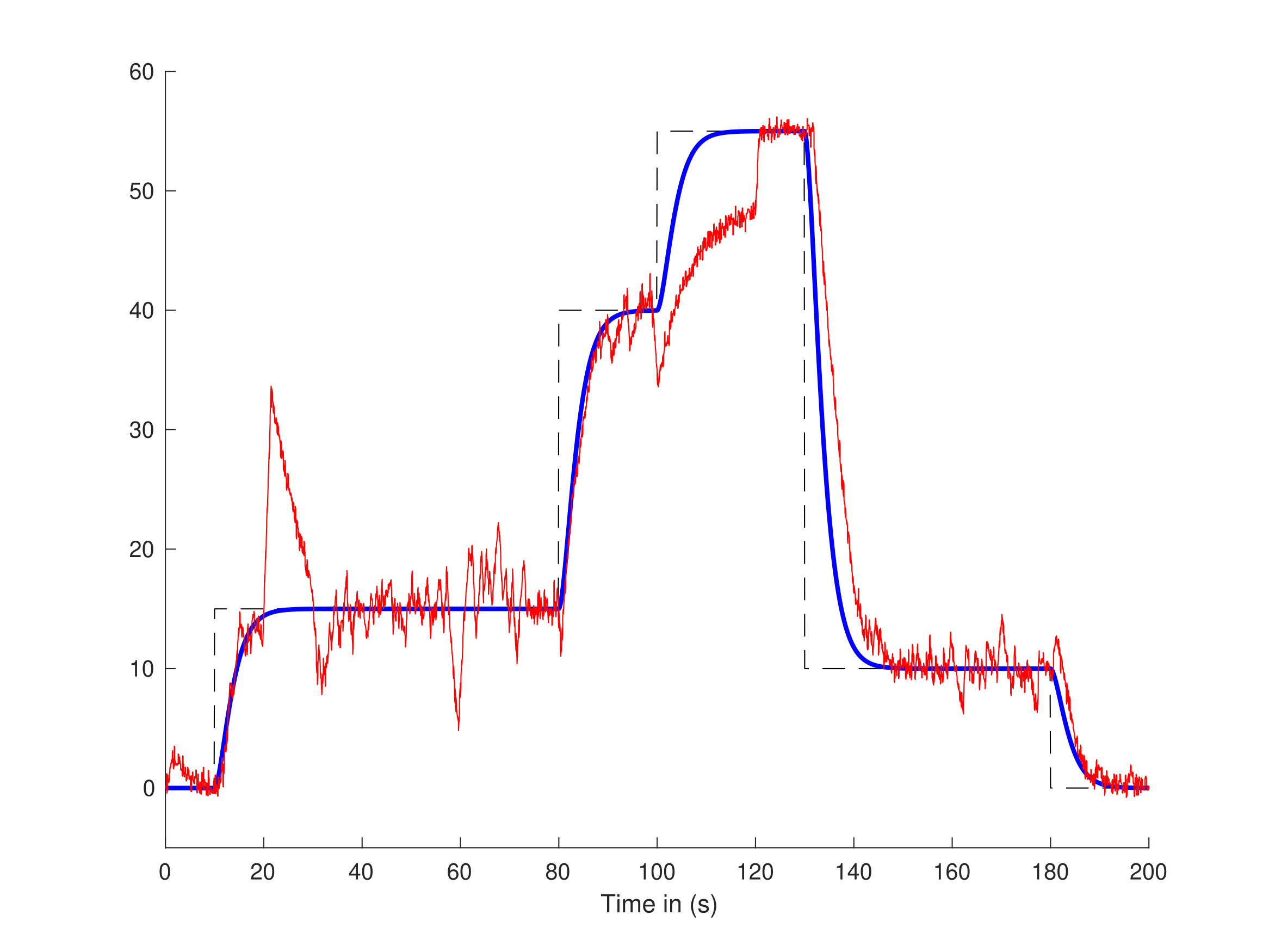,width=.2700\textwidth}}
\subfigure[\footnotesize Control variable]
{\epsfig{figure=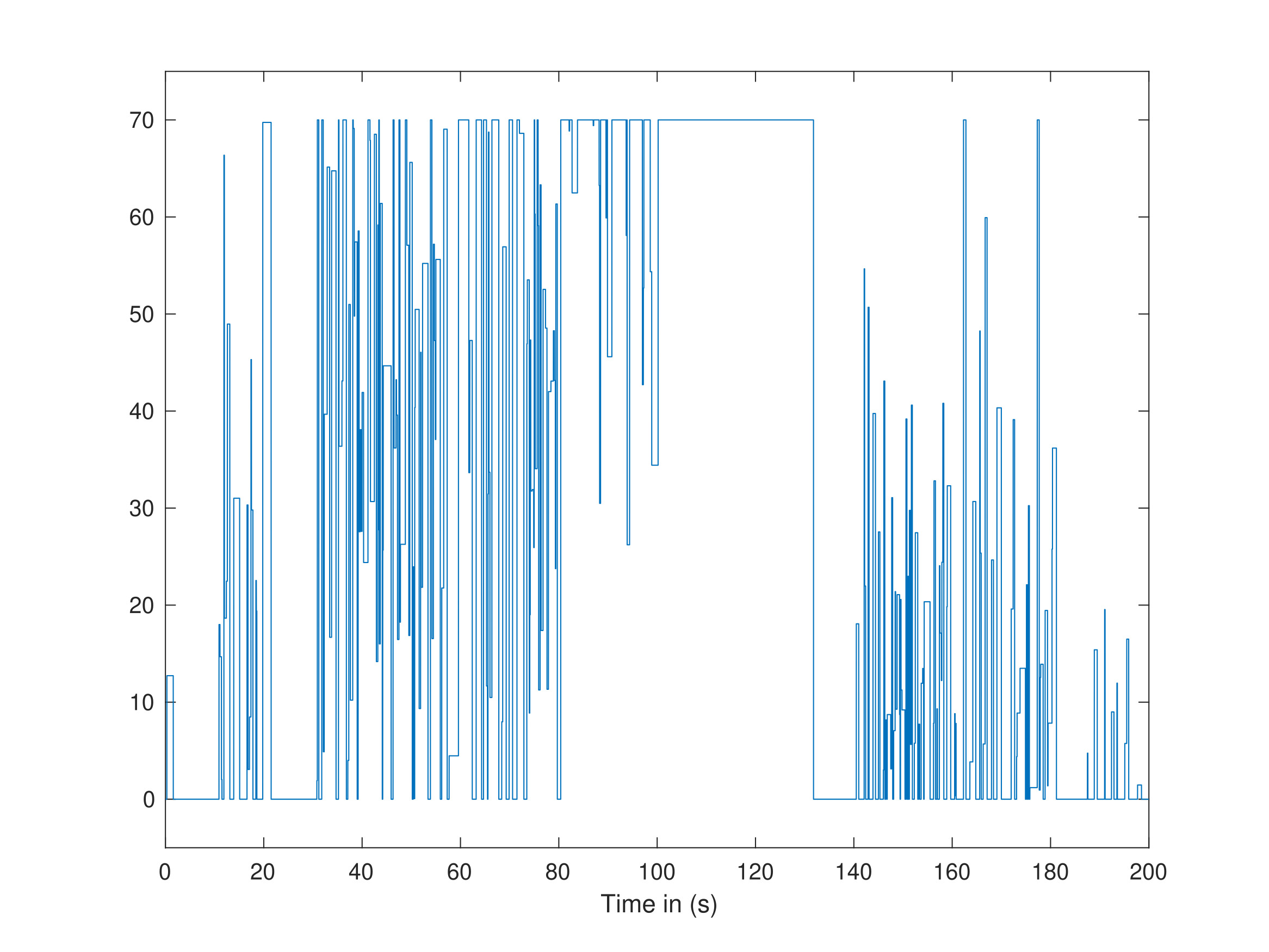,width=.2700\textwidth}}
\subfigure[\footnotesize Zoom on the faults]
{\epsfig{figure=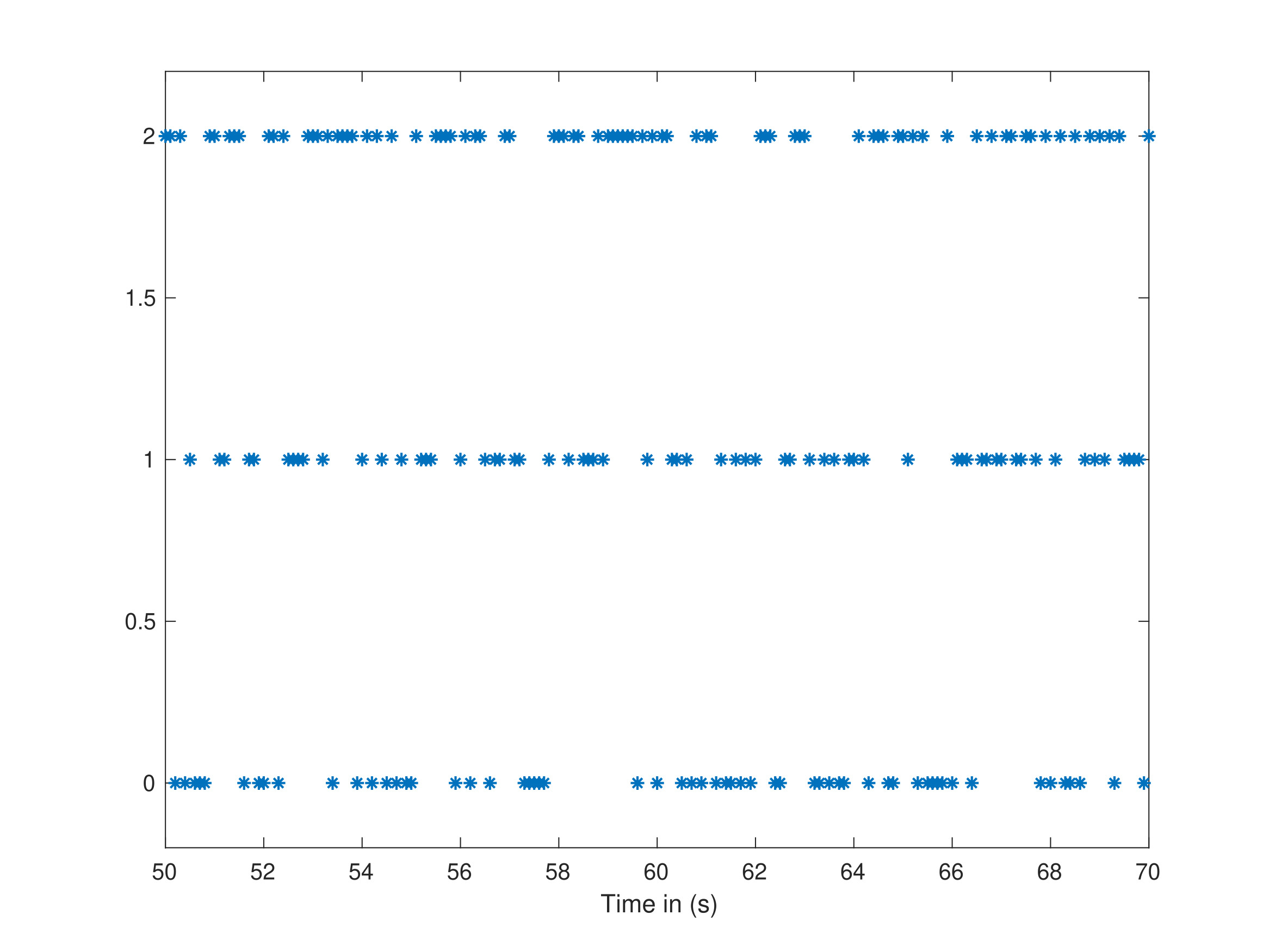,width=.2700\textwidth}}\\
\caption{Scénario 5 : PI }\label{Y5P}
\end{figure*}

Consider a classic PI controller (see, \textit{e.g.}, \cite{murray,broida}) where $e$ is the tracking error, $k_p, k_i \in \mathbb{R}$ are the gains:
\begin{equation}\label{pi}
u = k_p e + k_i \int e
\end{equation}
Set for the tank $\frak{K} = 30$ and for Formula \eqref{pi} $k_p=29.69$, $k_i=2.27009$.\footnote{Those numerical values are obtained via the Bro\"{\i}da method which is very popular in France (see, \textit{e.g.}, \cite{broida}).} The results in Figure \ref{Y1}-(c) are rather good without any packet loss, although $u$ (see Figure \ref{Y1}-(d)) is quite sensitive to the corrupting noise. When the packet loss become important Figure \ref{Y5P} shows a poor tracking. The malfunction depicted in Figure \ref{Y2P} is due to the usual \emph{anti-windup}, which is related to the integral term in Equation \eqref{pi} (see, \textit{e.g.}, \cite{murray,broida}).
\begin{remark}
In another situation, where a delay cannot be neglected, it has been shown \cite{stankovic1} that our iP behaves better than a classic PI.
\end{remark}

\section{Experiments with the Quanser AERO}\label{quanser}
\subsection{Quick presentation}
The Quanser AERO\footnote{See the link \newline {\tt https://www.quanser.com/products/quanser-aero/} \\ where a detailed picture is available.} is a half-quadrotor, which ``is a fully integrated dual-motor lab experiment, designed for advanced control research and aerospace applications.''  Two motors driving the propellers, which might turn clockwise or not, are controlling the angular position $y$ (rad) of the arms. Write $v_i$, $i = 1, 2$, the supply voltage of motor $i$, where $- 24{\rm v} \leq v_i \leq 24{\rm v}$ (volt).  


\subsection{Some experiments}\label{A}

\begin{figure*}[!ht]
\centering%
\subfigure[\footnotesize  Output (blue), reference trajectory (red)]
{\epsfig{figure=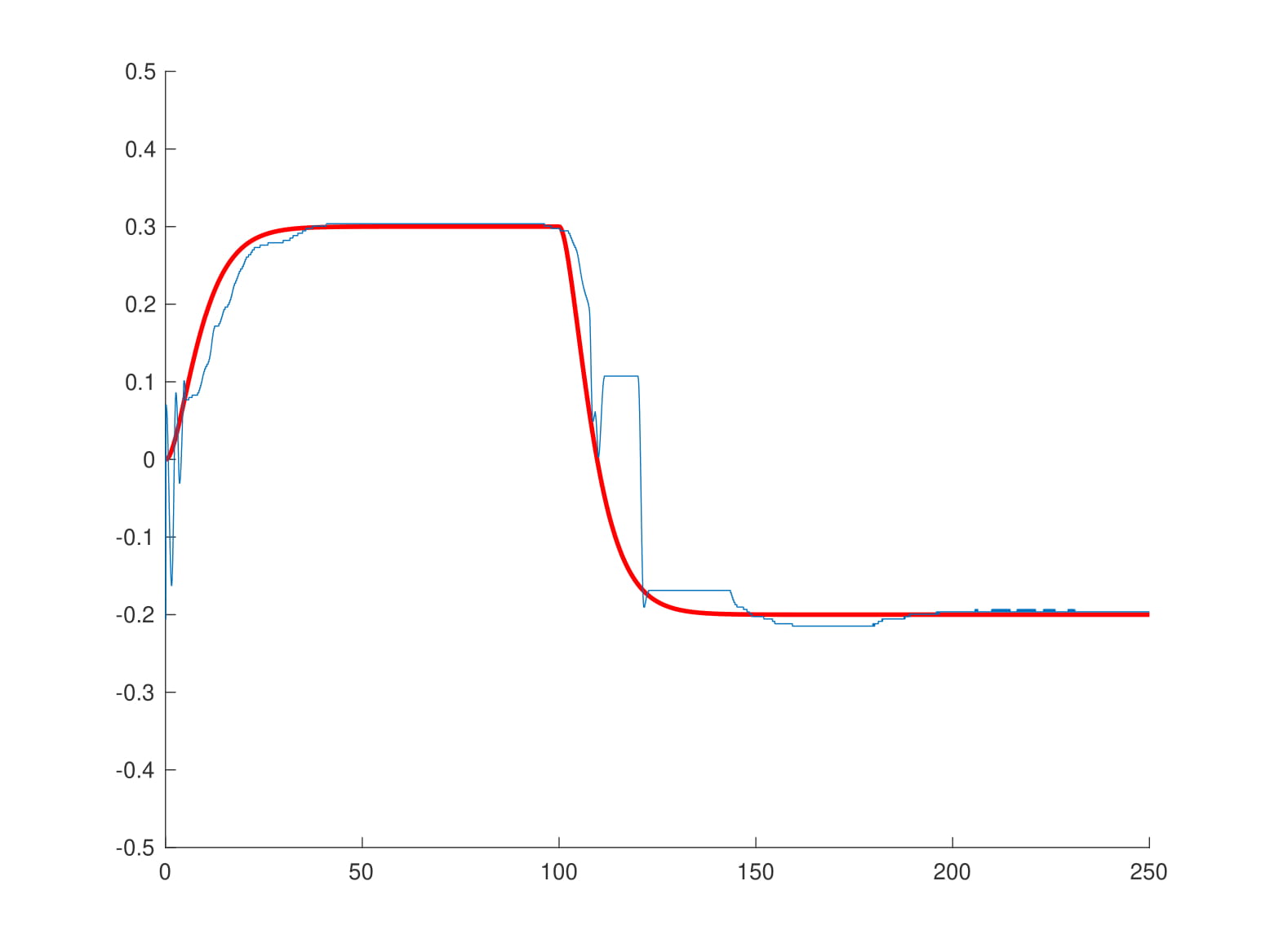,width=0.27002\textwidth}}
\subfigure[\footnotesize Supply voltages $v_1$ (blue), $v_2$ (red) ]
{\epsfig{figure=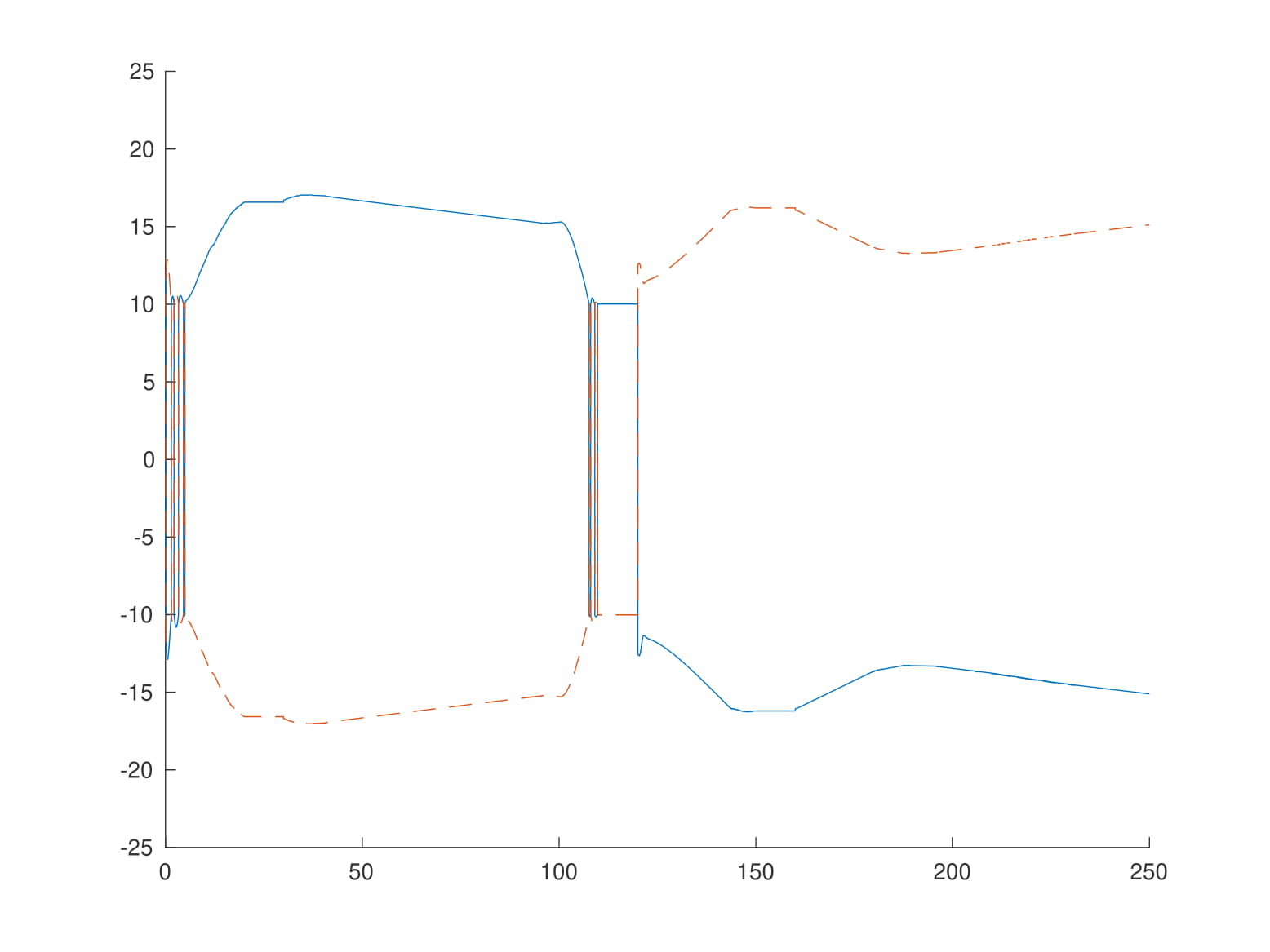,width=0.27002\textwidth}}
\subfigure[\footnotesize Various faults: \{0,1,2\} = {no fault, fault 1, fault 2} ]
{\epsfig{figure=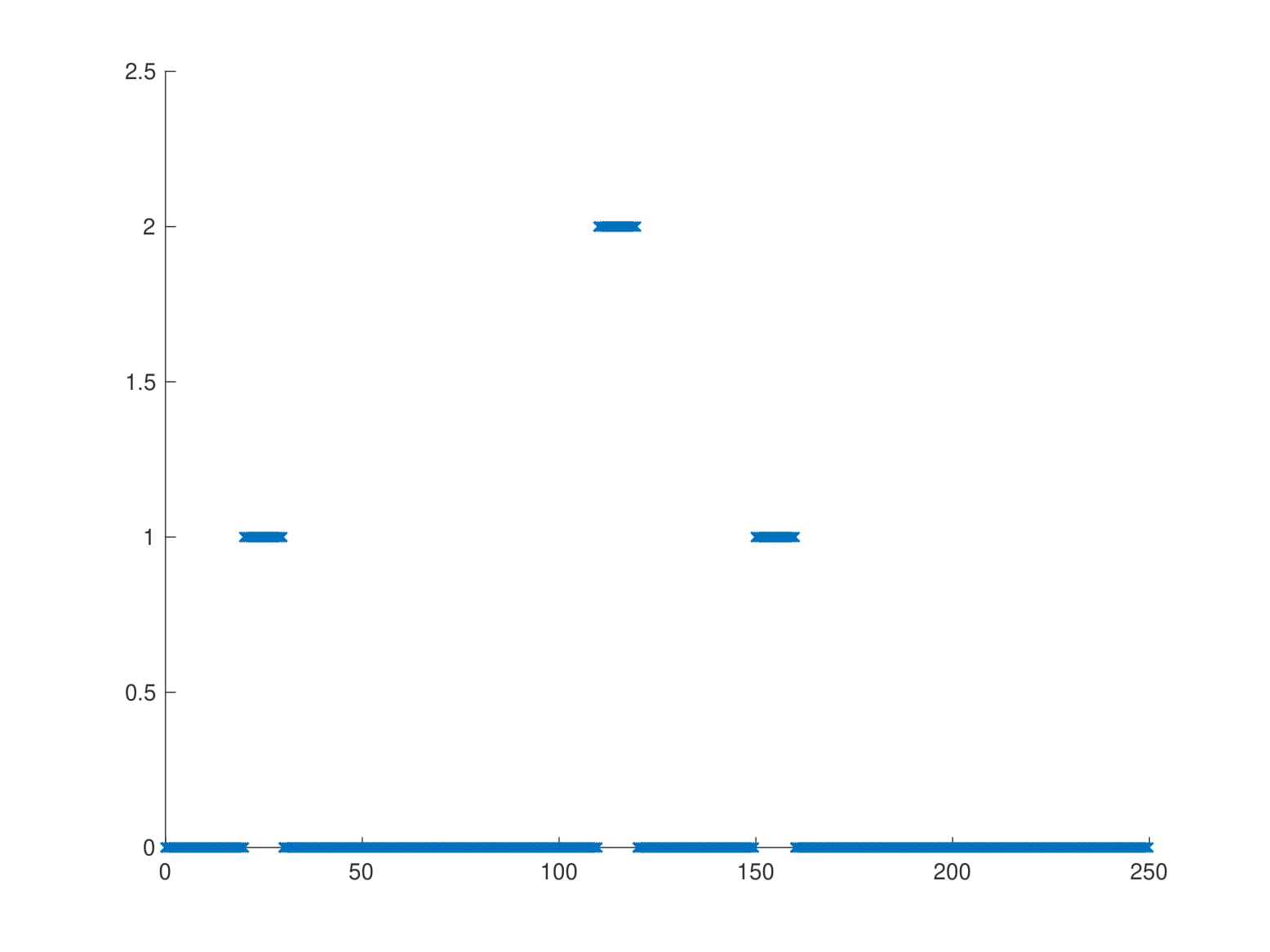,width=0.27002\textwidth}}
\caption{Scenario 1}\label{S8}
\end{figure*}

\begin{figure*}[!ht]
\centering%
\subfigure[\footnotesize Output (blue), reference trajectory (red)]
{\epsfig{figure=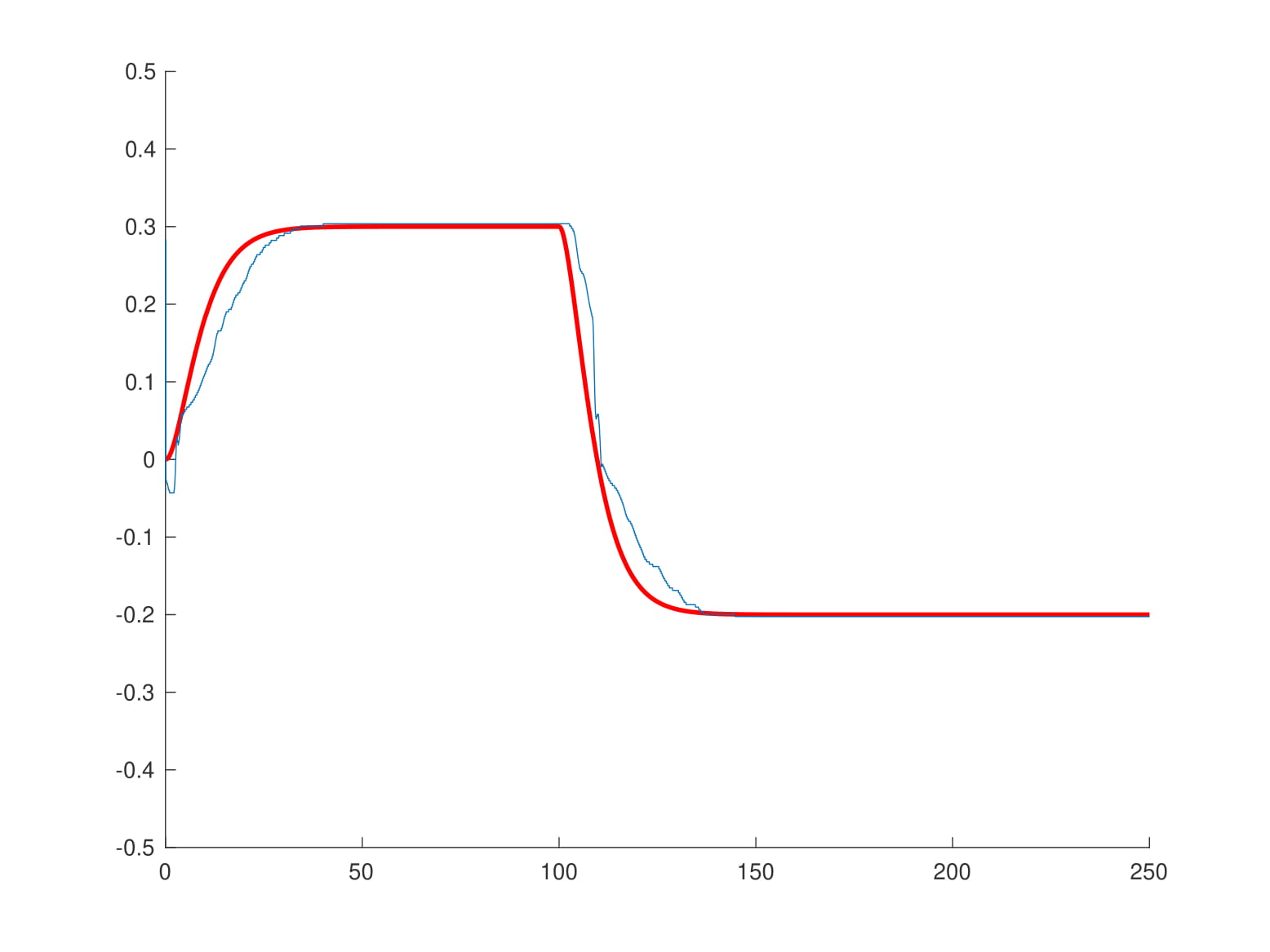,width=0.270\textwidth}}
\subfigure[\footnotesize Supply voltages $v_1$ (blue), $v_2$ (red) ]
{\epsfig{figure=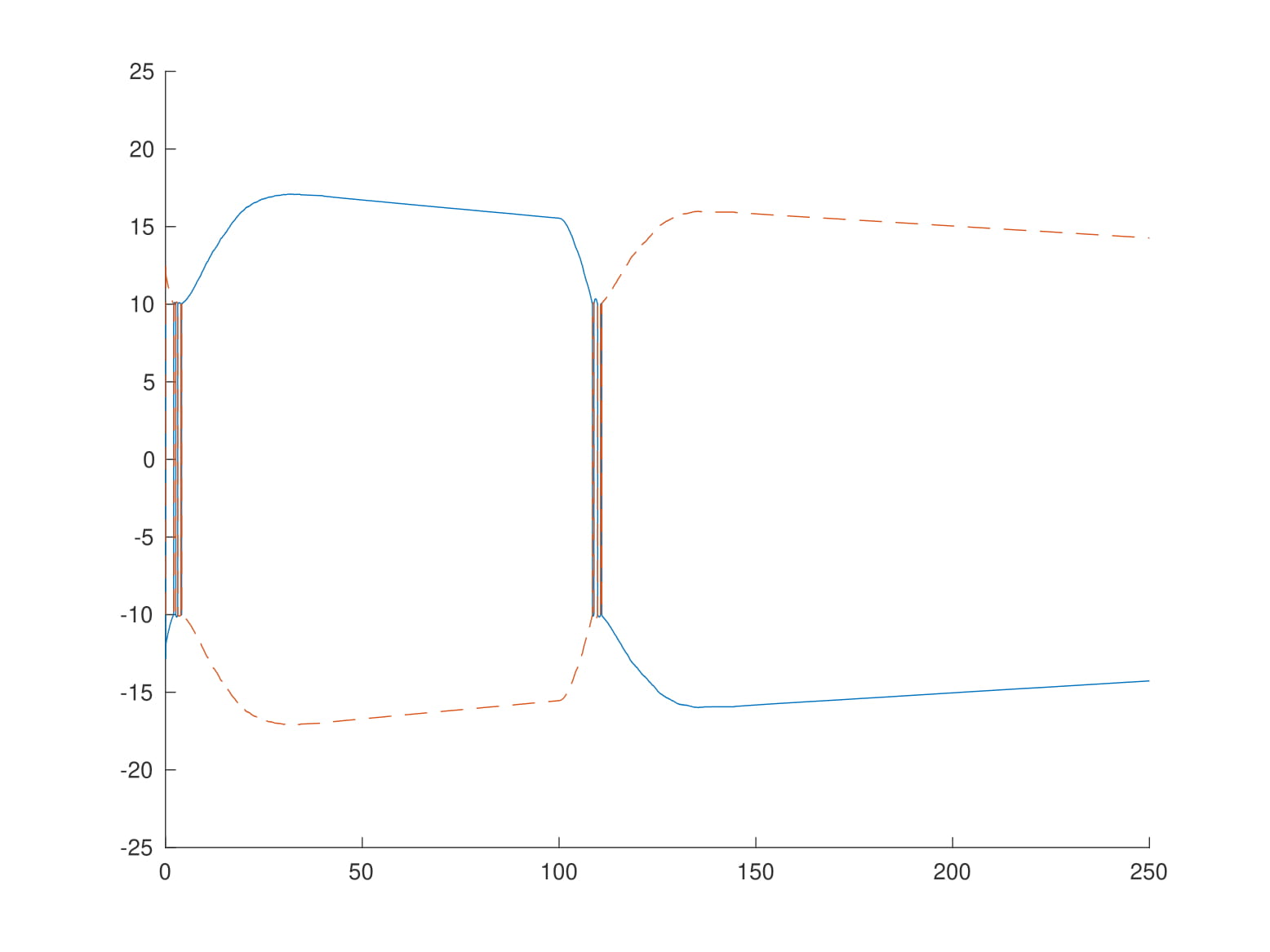,width=0.270\textwidth}}
\subfigure[\footnotesize Zoom on the faults: \{0,1,2\} = {no fault, fault 1, fault 2} ]
{\epsfig{figure=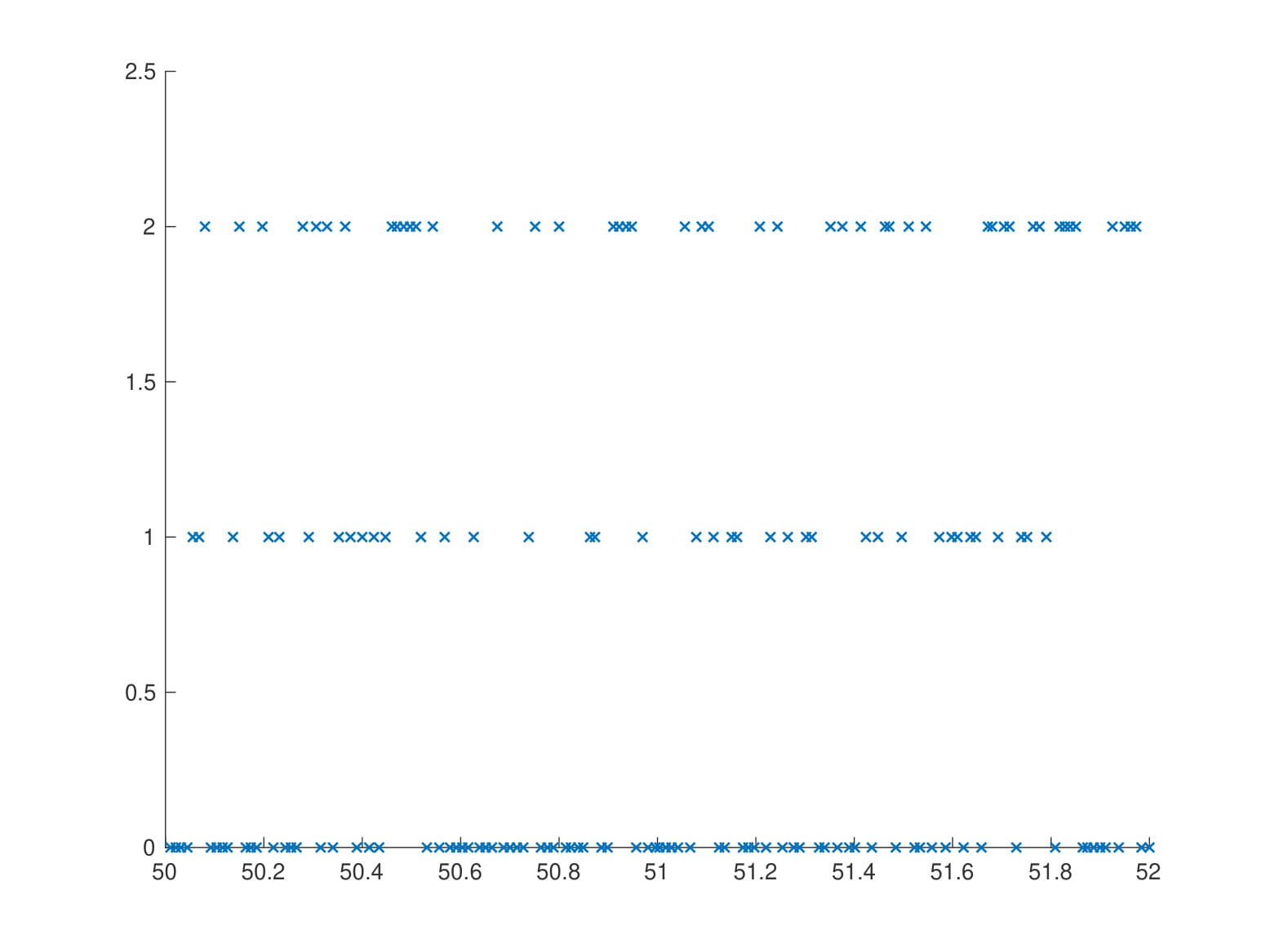,width=0.270\textwidth}}
\caption{Scenario 2}\label{S9}
\end{figure*}

\begin{figure*}[!ht]
\centering%
\subfigure[\footnotesize Output (blue), reference trajectory (red)]
{\epsfig{figure=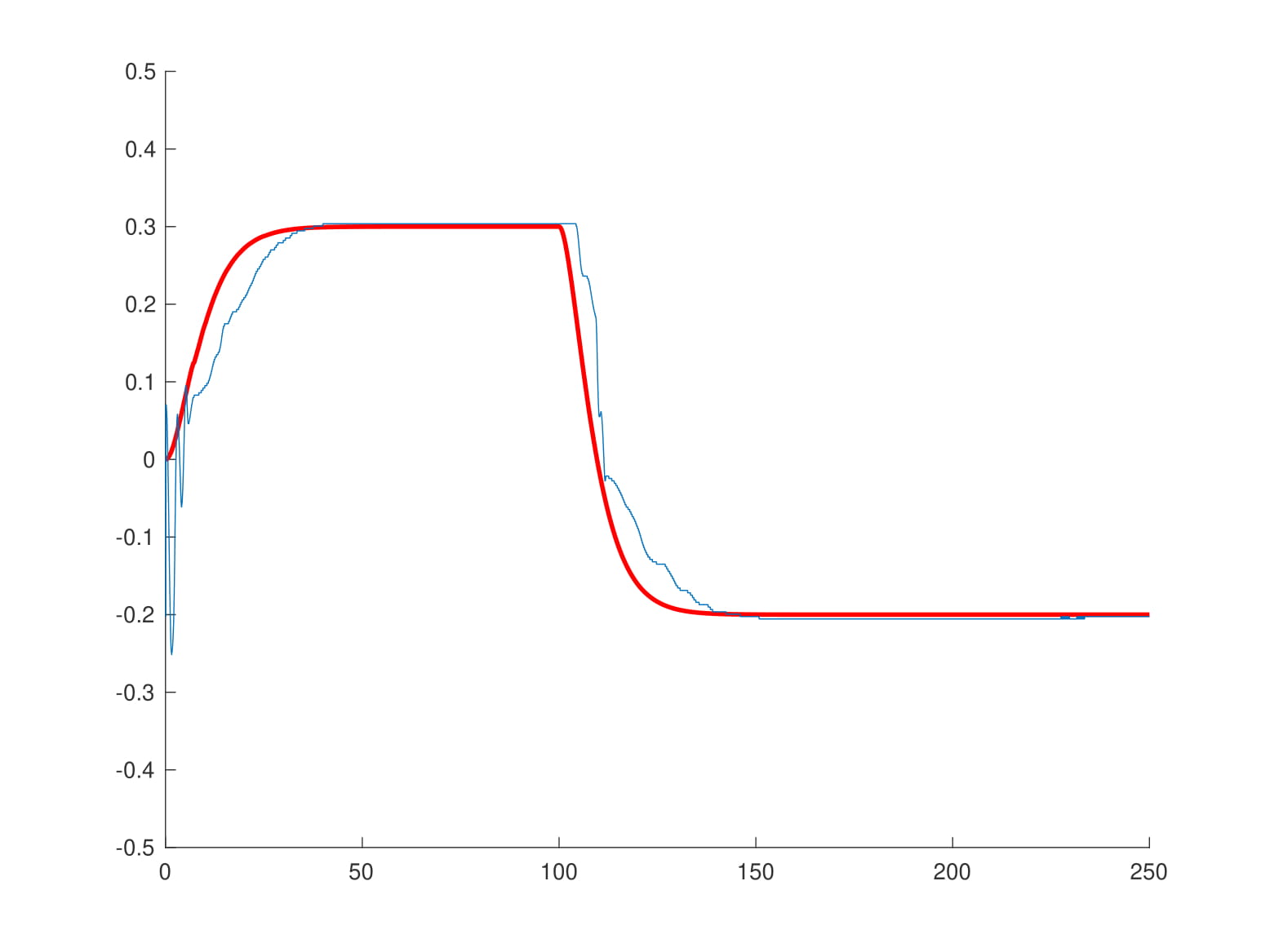,width=0.270\textwidth}}
\subfigure[\footnotesize Supply voltages $v_1$ (blue), $v_2$ (red) ]
{\epsfig{figure=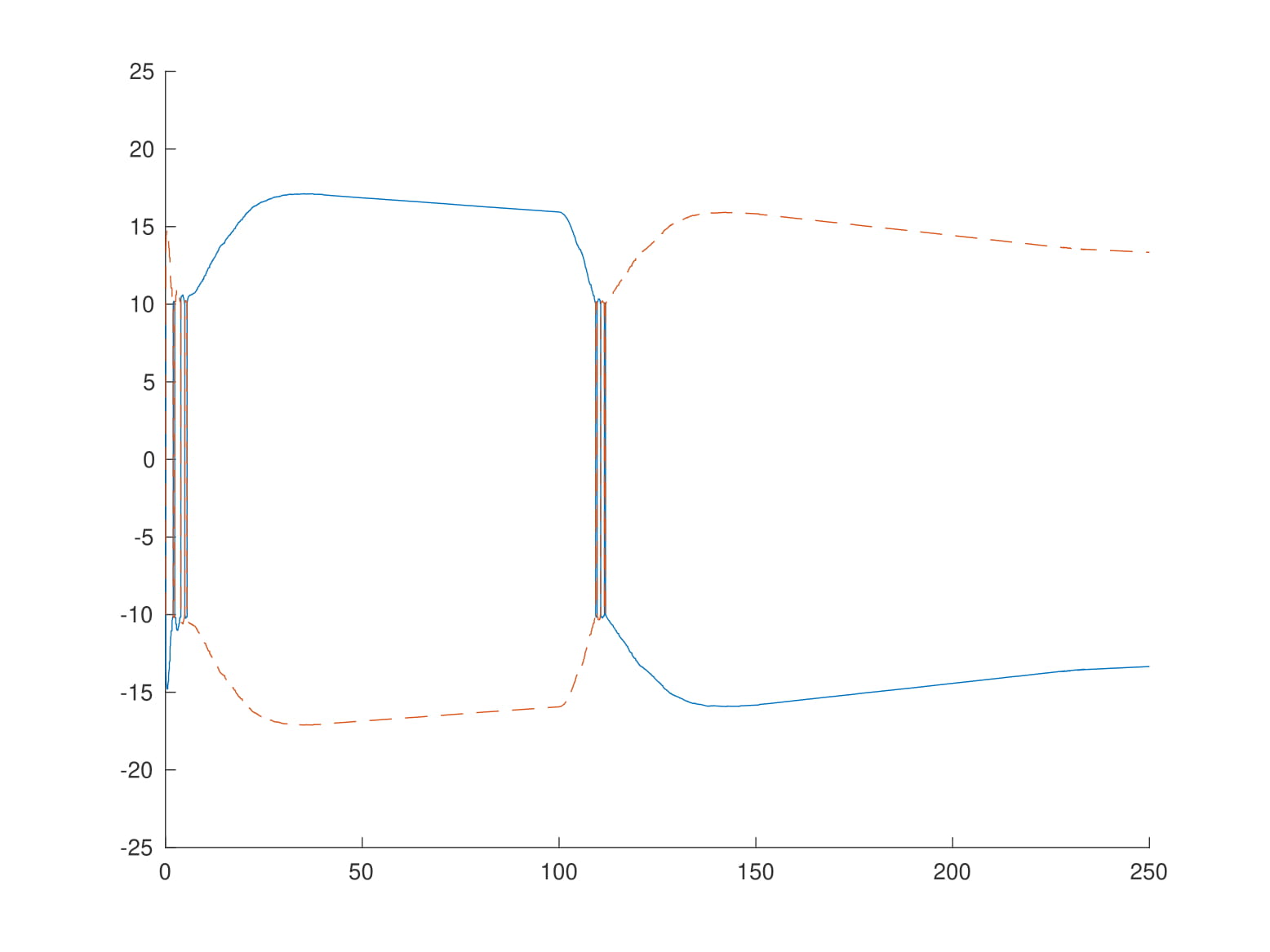,width=0.270\textwidth}}
\subfigure[\footnotesize Zoom on the faults: \{0,1,2\} = {no fault, fault 1, fault 2}  ]
{\epsfig{figure=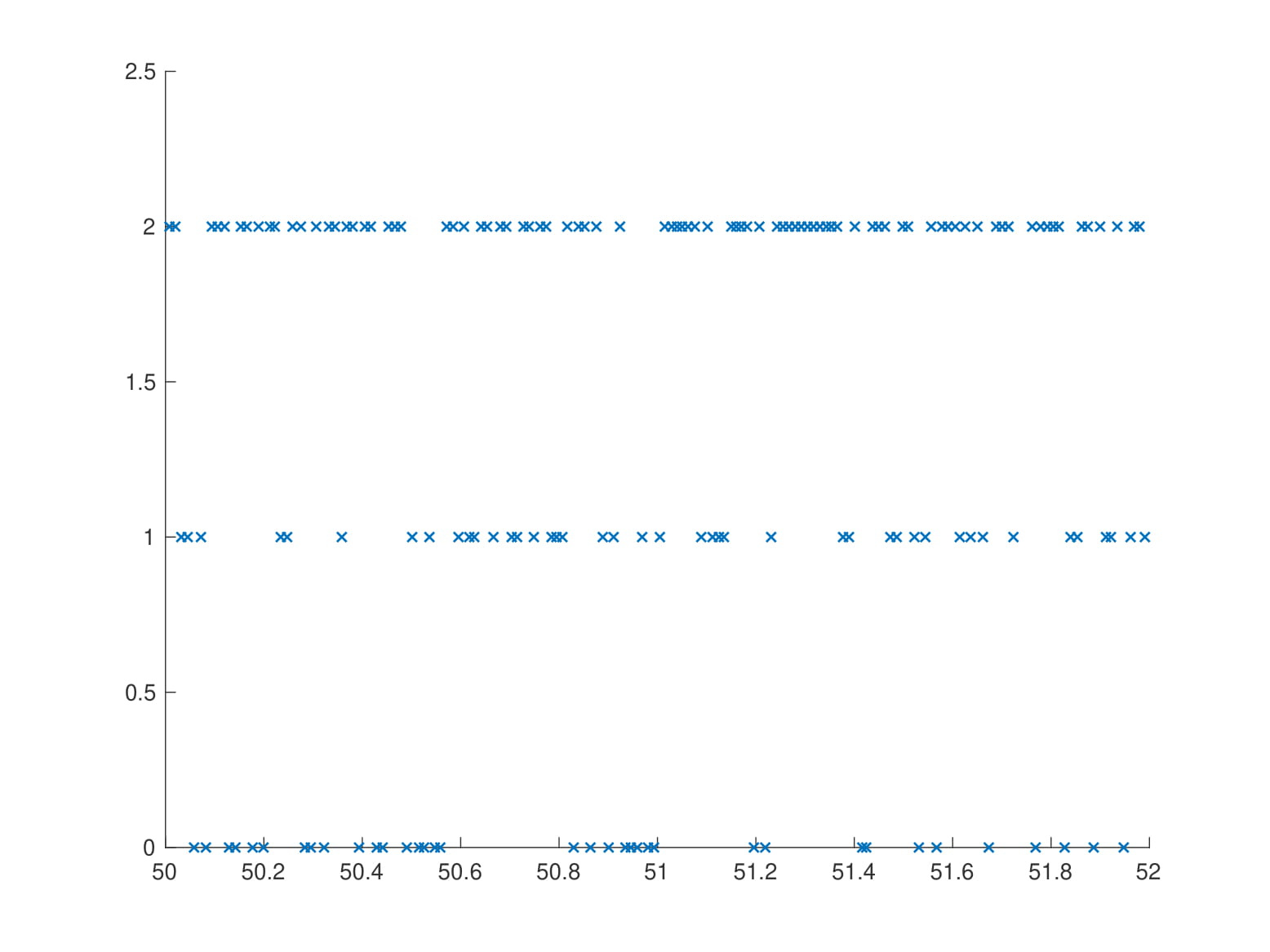,width=0.270\textwidth}}
\caption{Scenario 3}\label{S10}
\end{figure*}

The single control variable $u$ in Equation \eqref{1}  is defined by
\begin{itemize}
\item if $u>0$, then $v_1=10+u$, $v_2=-10-u$ 
\item if $u < 0$, then $v_1=-10+u$, $v_2=10-u$.
\end{itemize}
In Equations \eqref{1}-\eqref{ip} moreover, $\alpha=5$, $K_P=-10$.
Everything is programed in $C\#$ and stored in the server. It computes $u$ and $F_{\text{est}}$, every $10$ms, according to the process interface instructions. 
Consider again the types of packet loss of Section \ref{gen}.
The duration of the experiments is equal to $250$s. Three scenarios are examined:
\begin{itemize}
\item {\bf Scenario 1} -- 2 long transmission cuts with fault 1, and 1 with fault 2.  
\item {\bf Scenario 2} -- between the process interface and the server $24.02\%$ of faults 1 and $24.85 \%$ of faults 2. 
\item {\bf Scenario 3} -- between the process interface and the server $39.27004\%$ of faults 1 and $39.64\%$ of fault 2. 
\end{itemize}
Figure \ref{S8} shows a lower quality of the tracking with the long cuts in scenario 1. Note that when the cut is over, the tracking becomes again good. For the scenarios 2 and 3, Figures  \ref{S9} and \ref{S10} display excellent performances, in spite of the very high packet loss in scenario 3.

\subsection{Use of a joystick}\label{joystick}
\subsubsection{The joystick}
A joystick $G_{\rm joystick}$ is assumed to impose a motion to the AERO. According to the ``philosophy'' of \emph{flatness-based} control (see \cite{flmr} and \cite{murray})
\begin{itemize}
\item it means to select thanks to the joystick an appropriate reference trajectory,
\item the iP \eqref{ip} ensures a good tracking. 
\end{itemize} 
Let us assume for simplicity's sake that this trajectory is deduced from the joystick's motion ${\text{Mot}} (G_{\rm joystick})$ via a linear filter with \emph{transfer function} (see, \textit{e.g.}, \cite{murray}) 
$$
\frac{1}{(Ts+1)^2}
$$ 
\begin{remark}
The relationship with \emph{cloud gaming} (see, \textit{e.g.}, \cite{claypool}) and \emph{telesurgery}  (see, \textit{e.g.}, \cite{korte}) is obvious.
\end{remark}
\subsubsection{Scenarios without any packet loss}
\begin{figure*}[!ht]
\centering%
\subfigure[\footnotesize Output (blue), reference trajectory (red), joystick (green)]
{\epsfig{figure=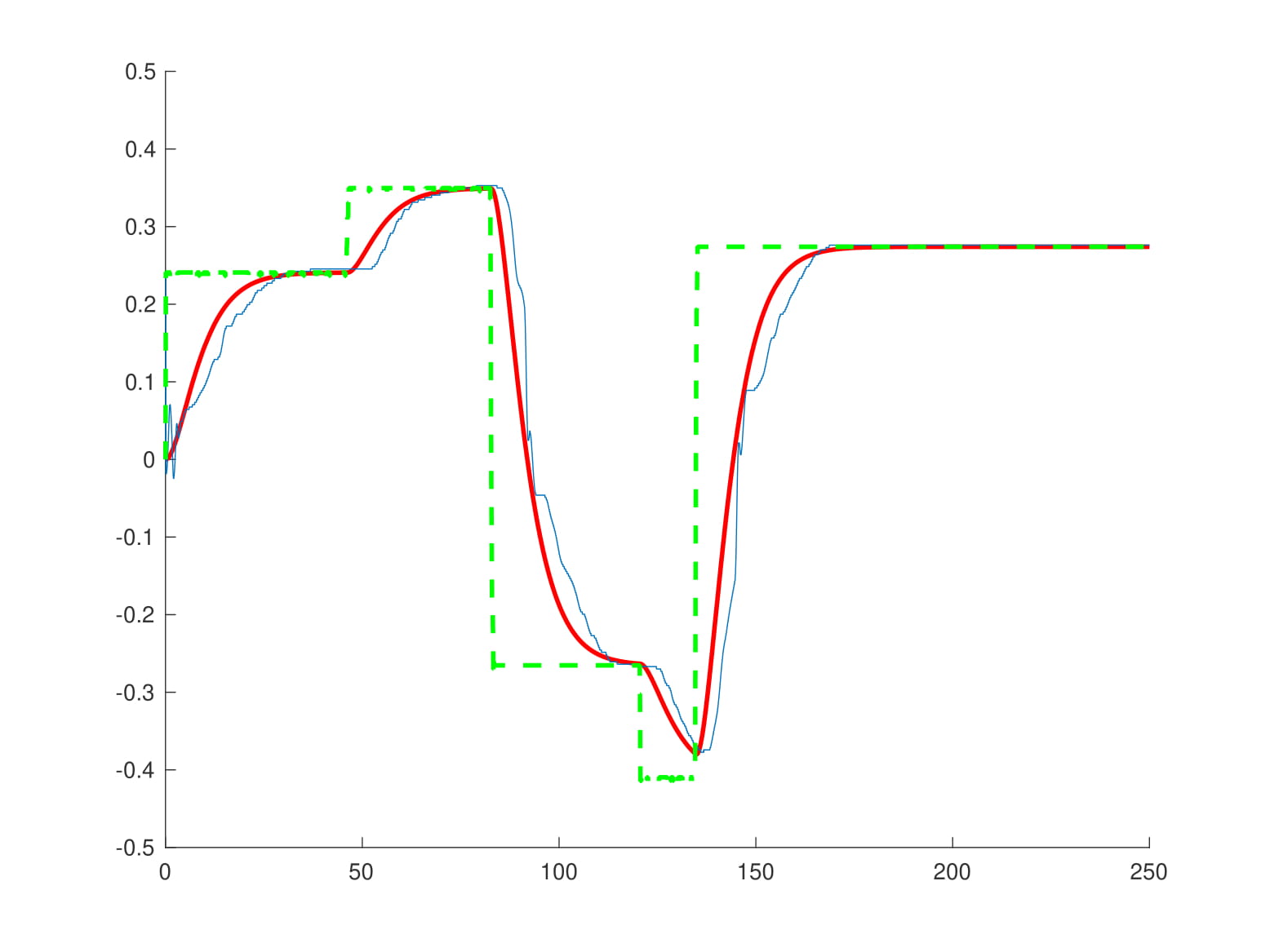,width=.50\textwidth}}
\subfigure[\footnotesize Supply voltages $v_1$ (blue), $v_2$ (red) ]
{\epsfig{figure=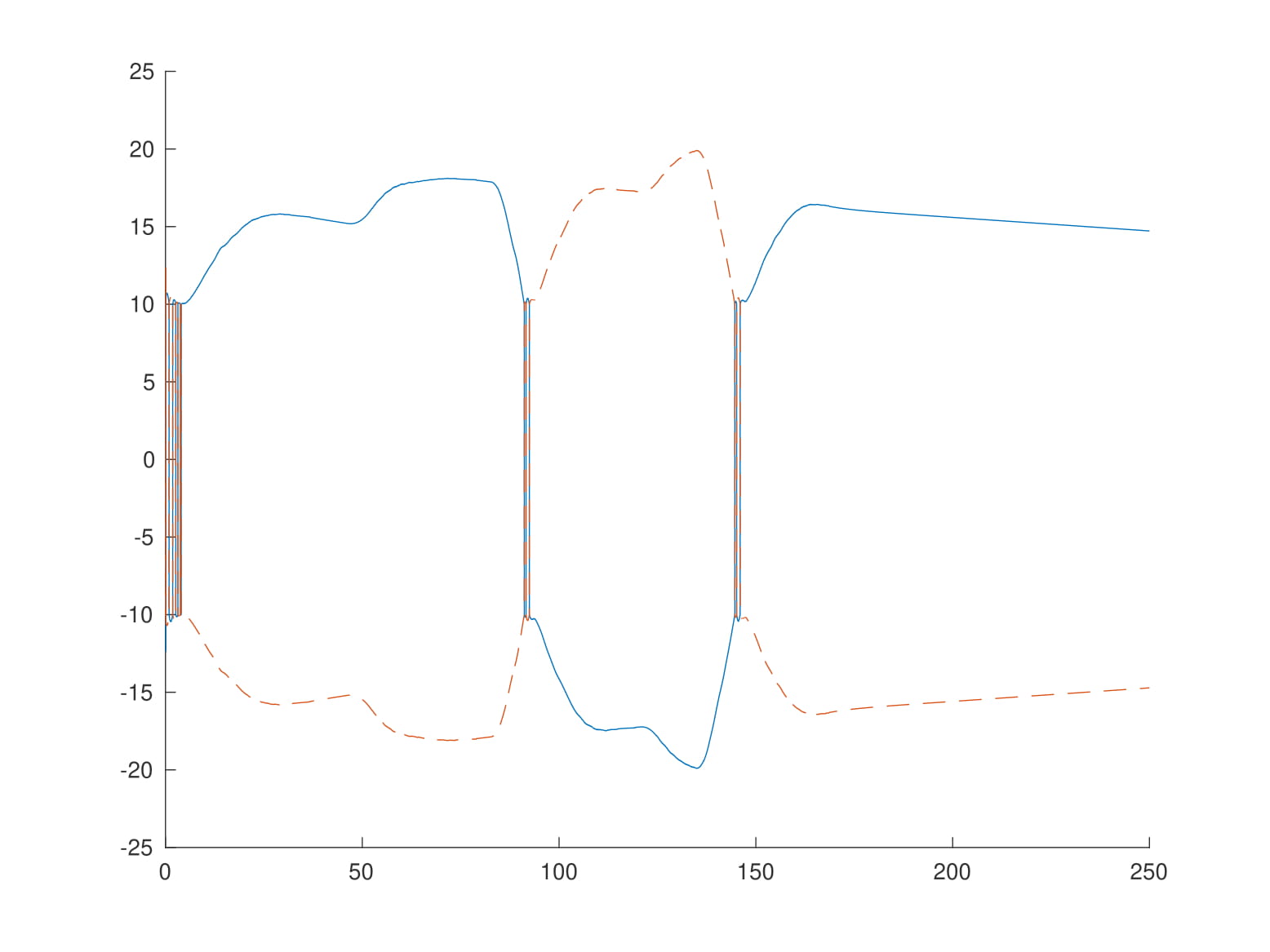,width=.50\textwidth}}
\caption{Scenario 4}\label{S11}
\end{figure*}

\begin{figure*}[!ht]
\centering%
\subfigure[\footnotesize Output (blue), reference trajectory (red), joystick (green)]
{\epsfig{figure=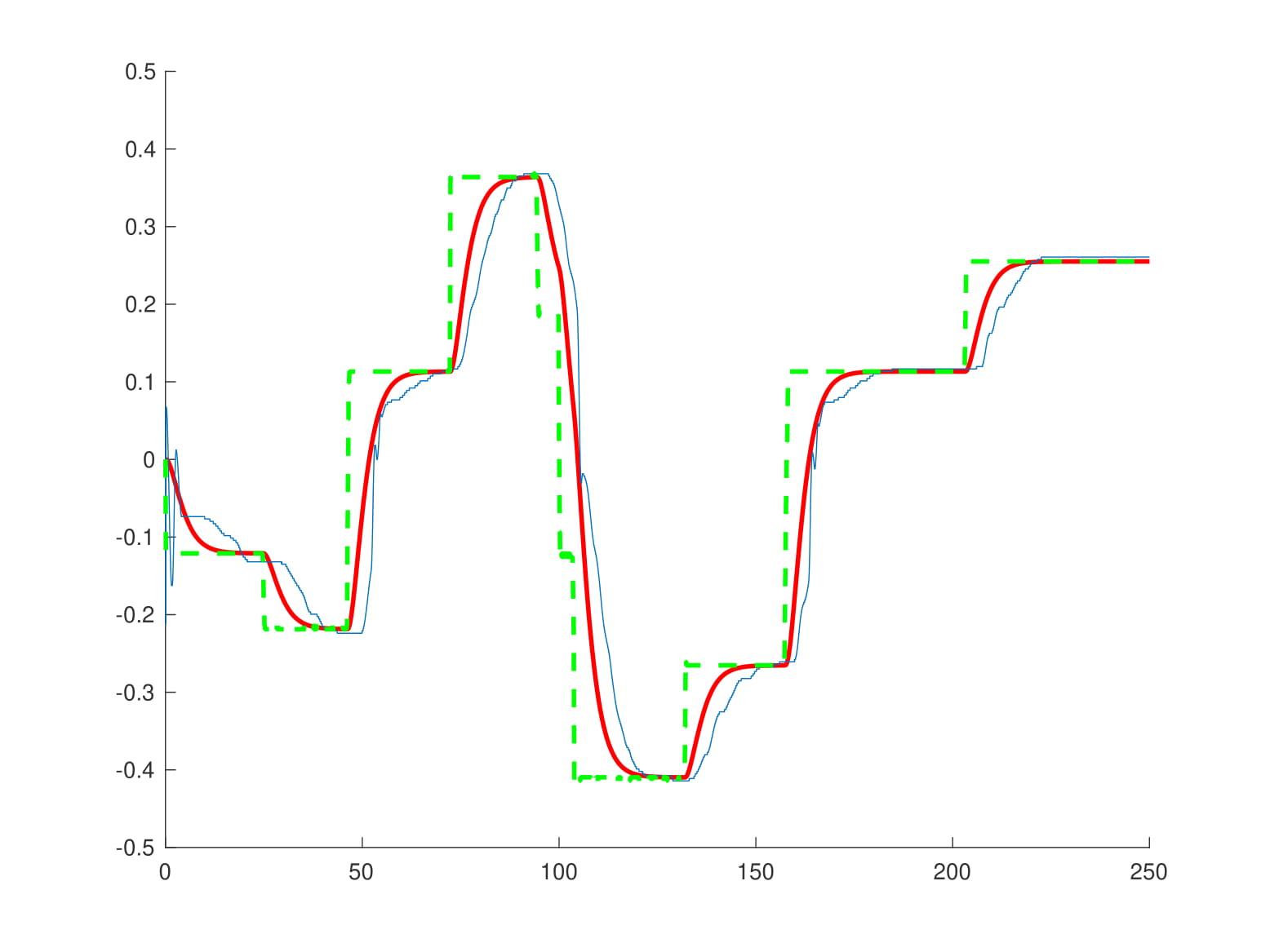,width=.50\textwidth}}
\subfigure[\footnotesize Supply voltages $v_1$ (blue), $v_2$ (red) ]
{\epsfig{figure=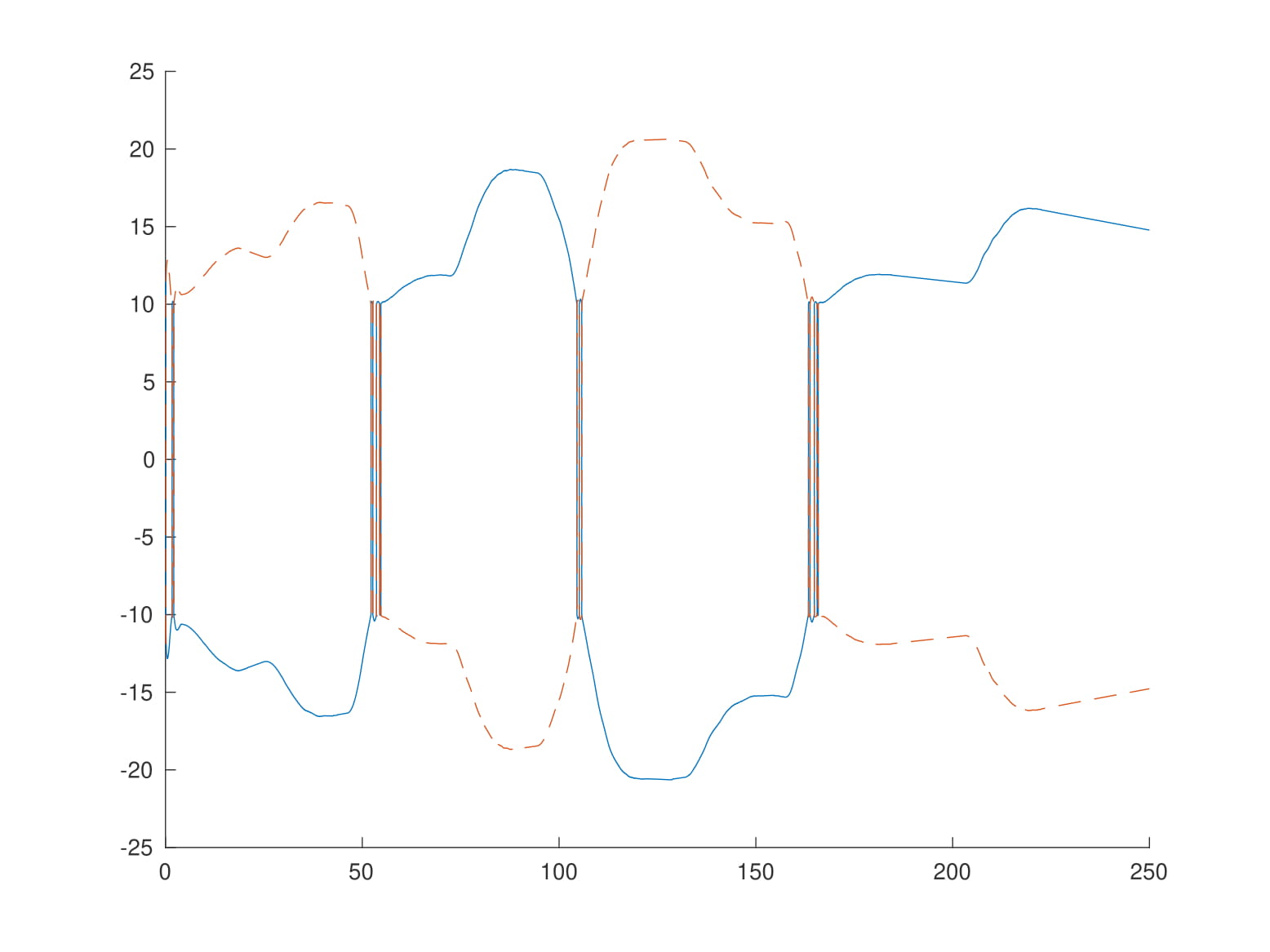,width=.50\textwidth}}
\caption{Scenario 5}\label{S12}
\end{figure*}

\begin{figure*}[!ht]
\centering%
\subfigure[\footnotesize Output (blue), reference trajectory (red), joystick (green)]
{\epsfig{figure=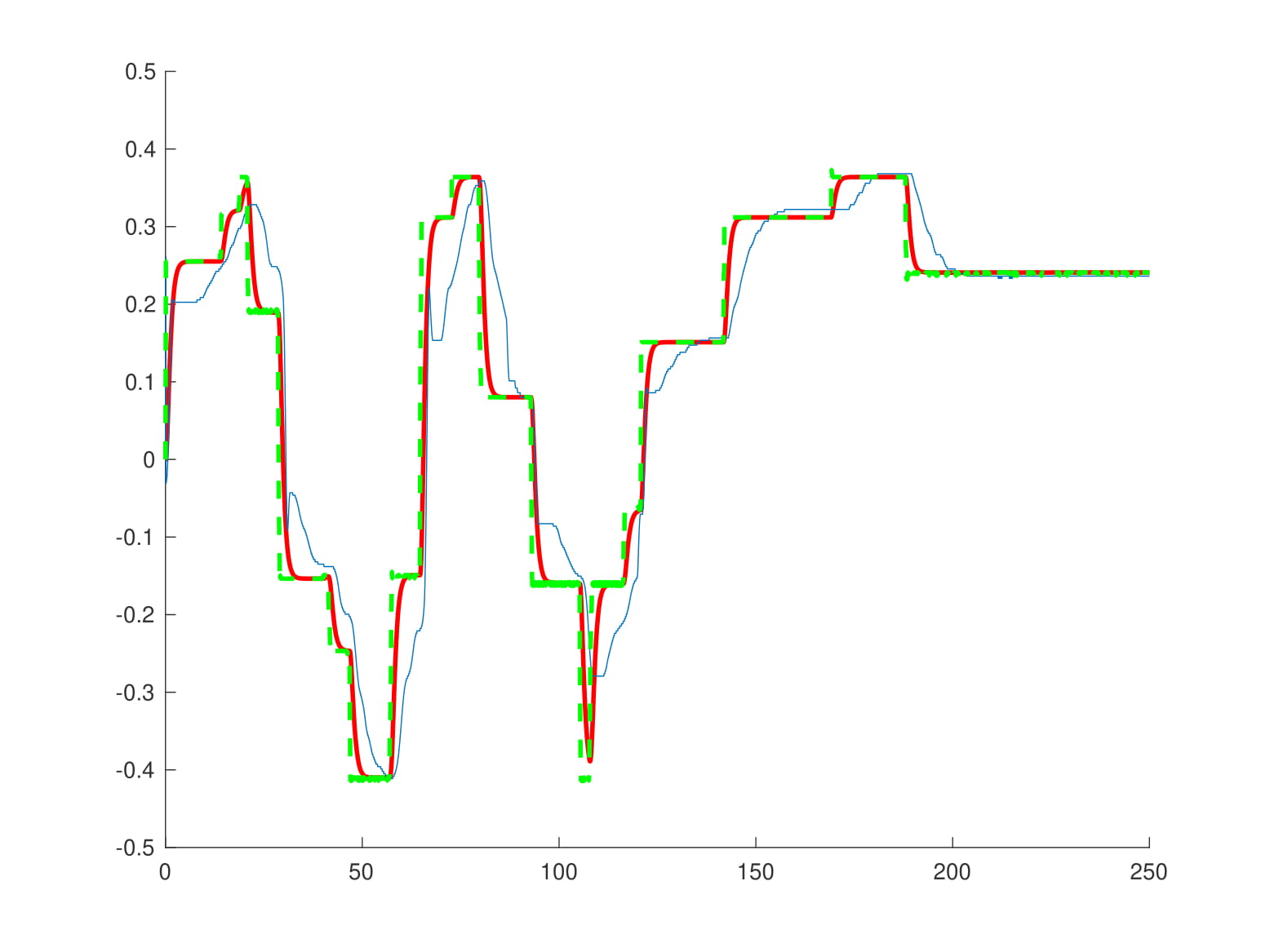,width=.50\textwidth}}
\subfigure[\footnotesize Supply voltages $v_1$ (blue), $v_2$ (red) ]
{\epsfig{figure=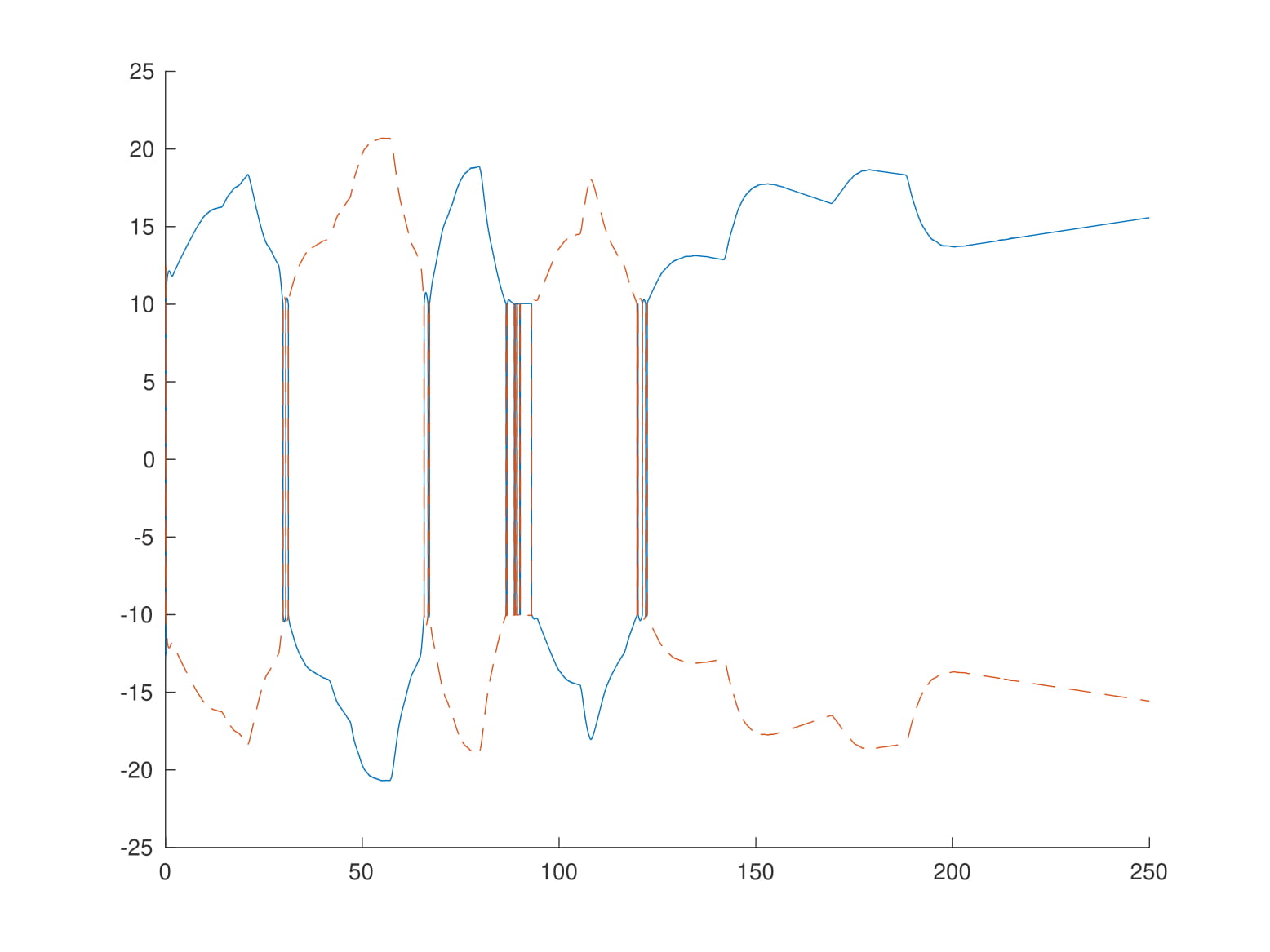,width=.50\textwidth}}
\caption{Scenario 6}\label{S13}
\end{figure*}

Three scenarios are again considered:
\begin{itemize}
\item {\bf Scenario 4} -- $T = 4$s.
\item {\bf Scenario 5} -- $T = 2$s.
\item {\bf Scenario 6} -- $T = 0.5$s.
\end{itemize}
Figures \ref{S11} and  \ref{S12} display an excellent tracking for the scenarios 4 and 5. A deterioration appears in  Figure \ref{S13} with respect to the scenario 6: this scenario is not really feasible from a purely mechanical viewpoint. Scenario 5 seems to be the best compromise between the speed of reaction and reachable trajectories. 
\subsubsection{Scenarios with packet loss}\label{joy}
\begin{figure*}[!ht]
\centering%
\subfigure[\footnotesize Output (blue), reference trajectory (red), joystick (green)]
{\epsfig{figure=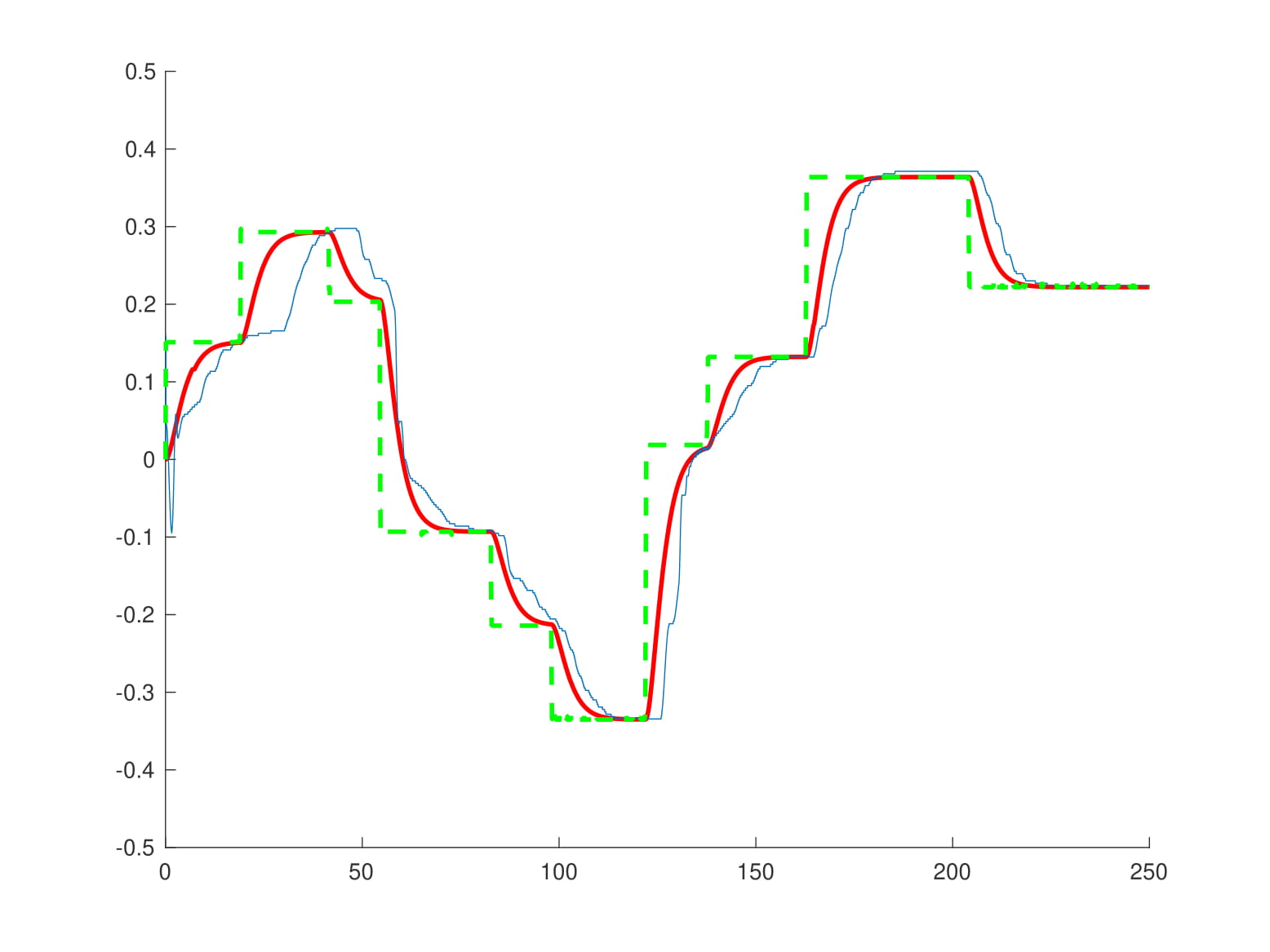,width=0.270\textwidth}}
\subfigure[\footnotesize Supply voltages $v_1$ (blue), $v_2$ (red) ]
{\epsfig{figure=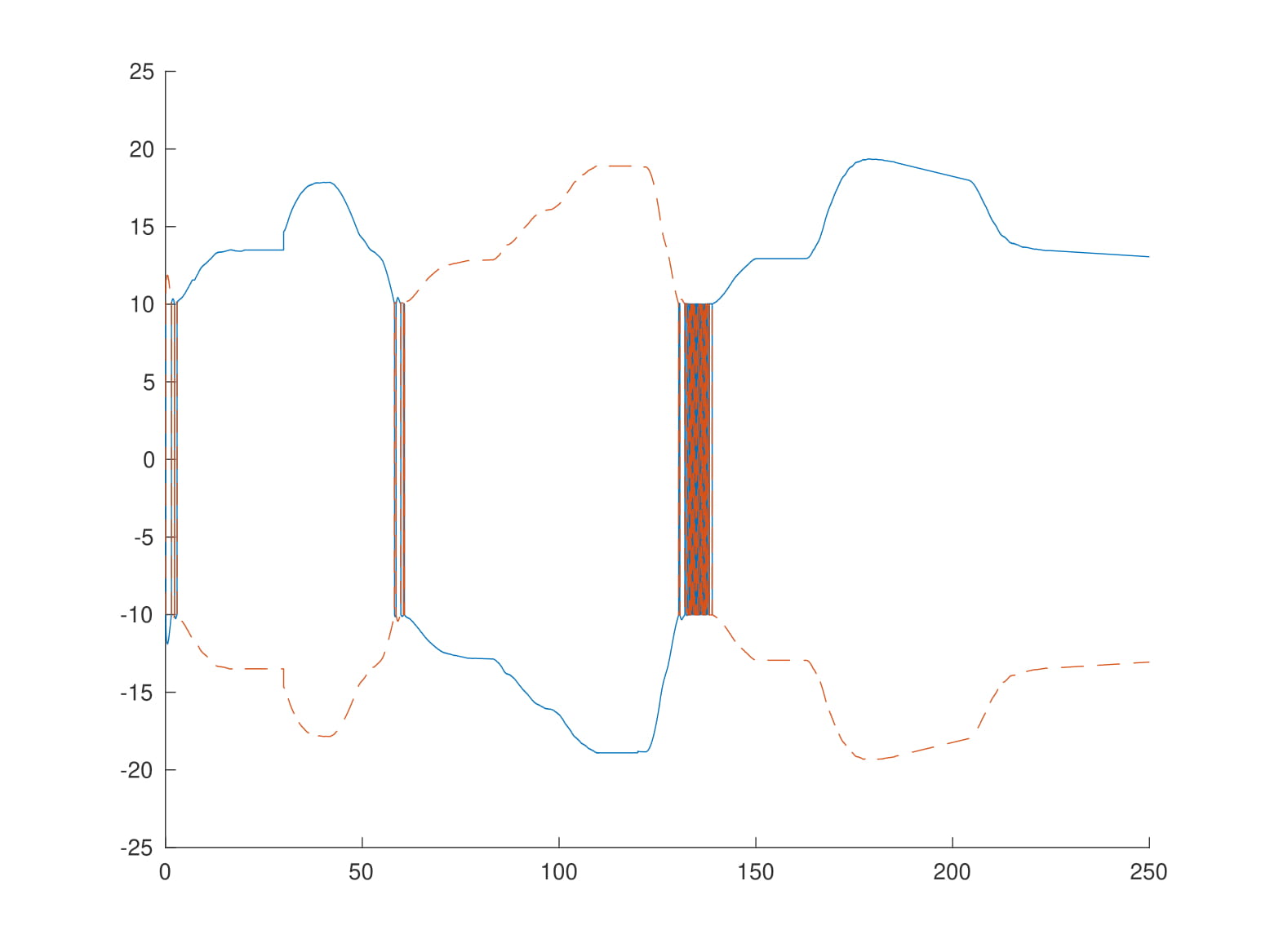,width=0.270\textwidth}}
\subfigure[\footnotesize Faults: \{0,1,2\} = \{no fault, fault 1, fault 2\} ]
{\epsfig{figure=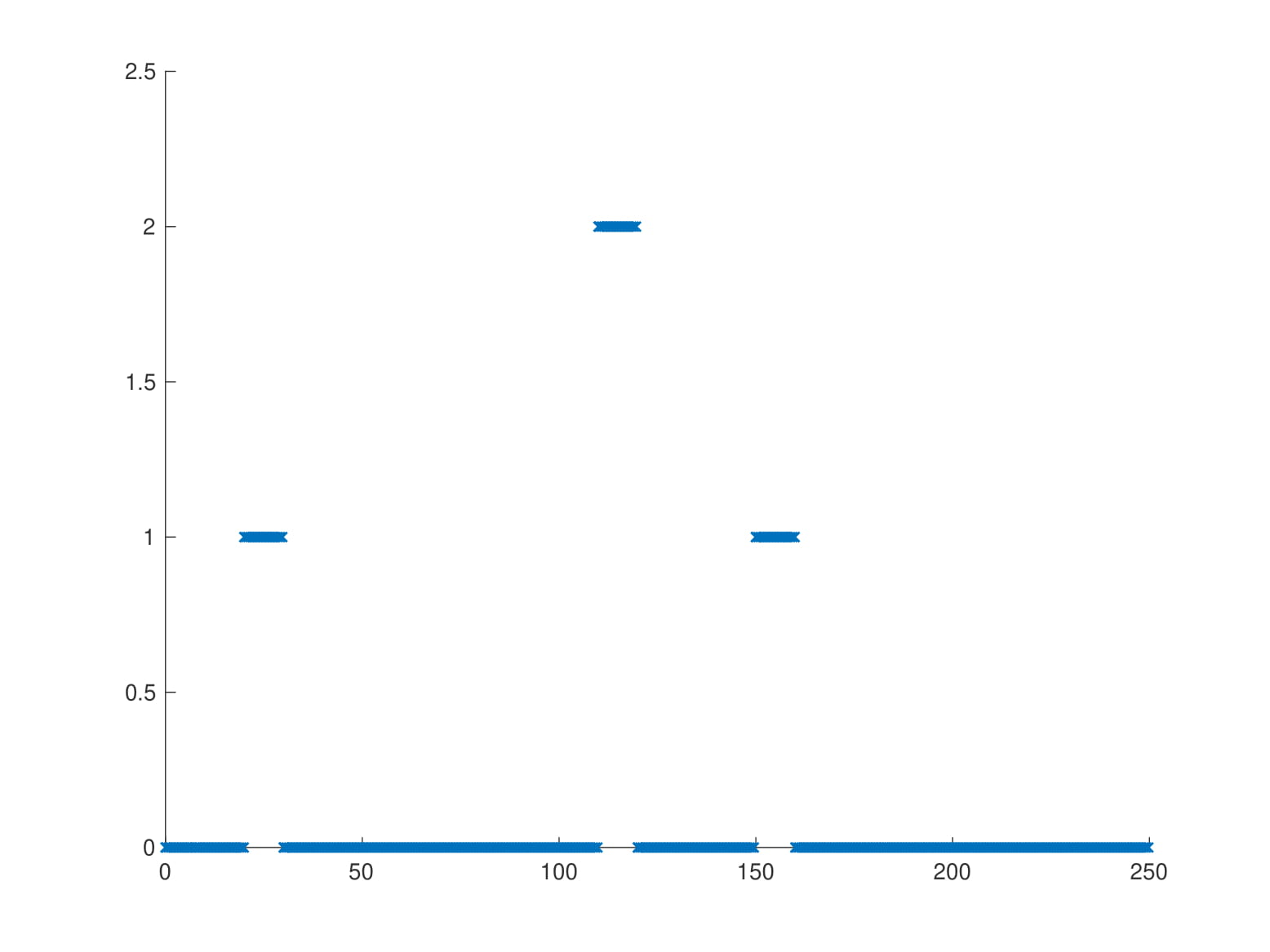,width=0.270\textwidth}}
\caption{Scenario 7}\label{S14}
\end{figure*}

\begin{figure*}[!ht]
\centering%
\subfigure[\footnotesize Output (blue), reference trajectory (red), joystick (green)]
{\epsfig{figure=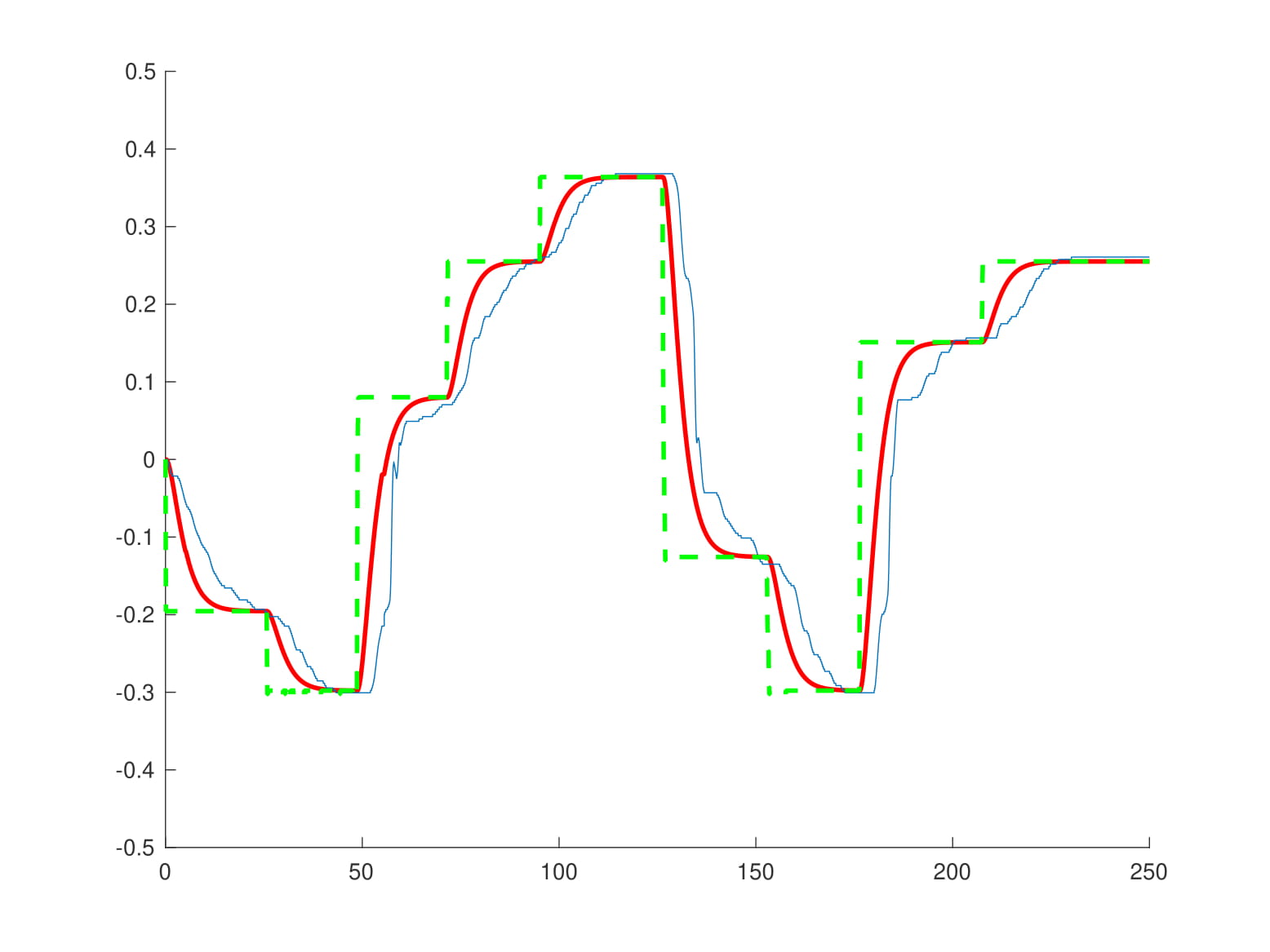,width=0.270\textwidth}}
\subfigure[\footnotesize Supply voltages $v_1$ (blue), $v_2$ (red) ]
{\epsfig{figure=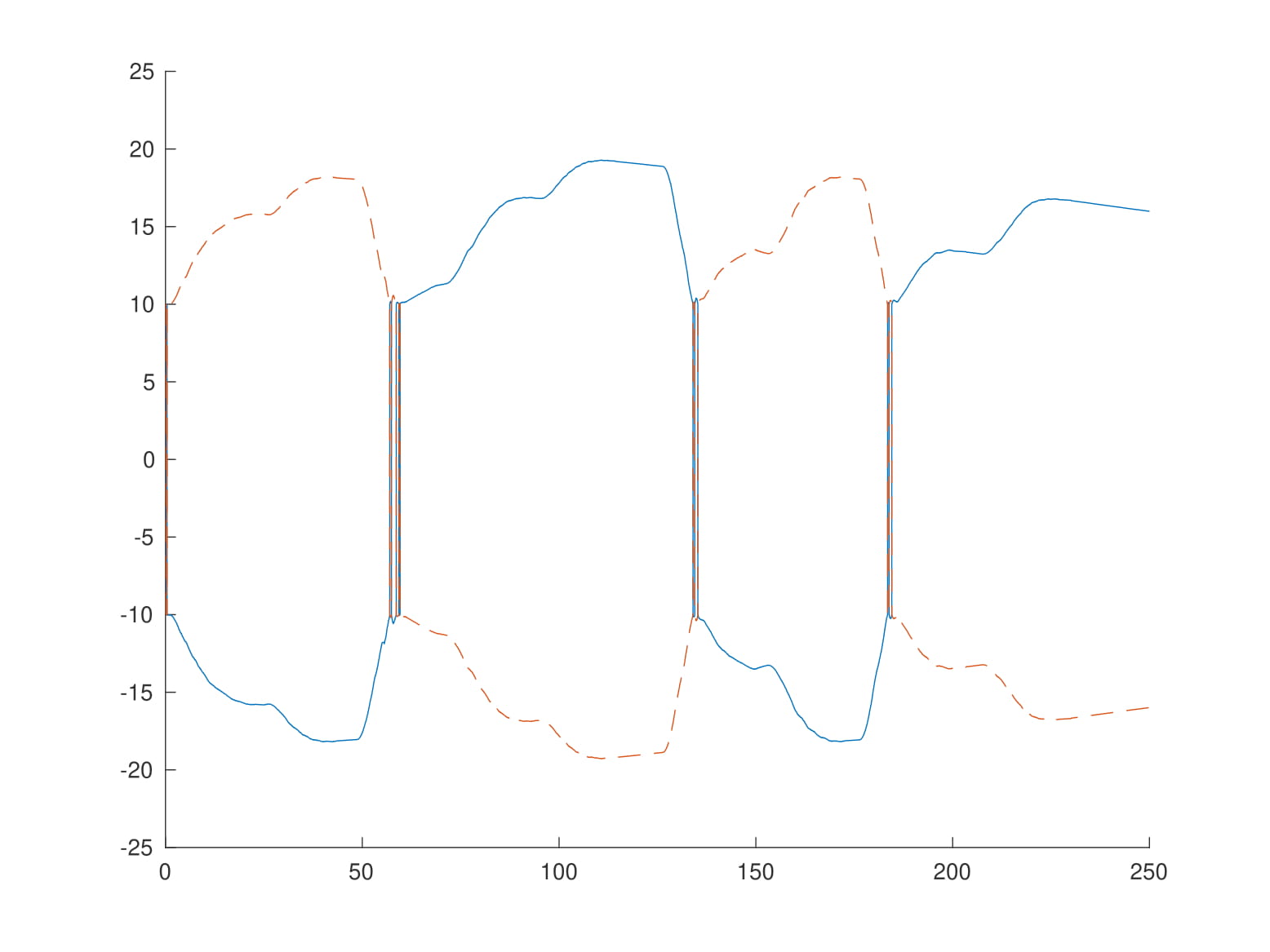,width=0.270\textwidth}}
\subfigure[\footnotesize Zoom on the faults: \{0,1,2\} = \{no fault, fault 1, fault 2\} ]
{\epsfig{figure=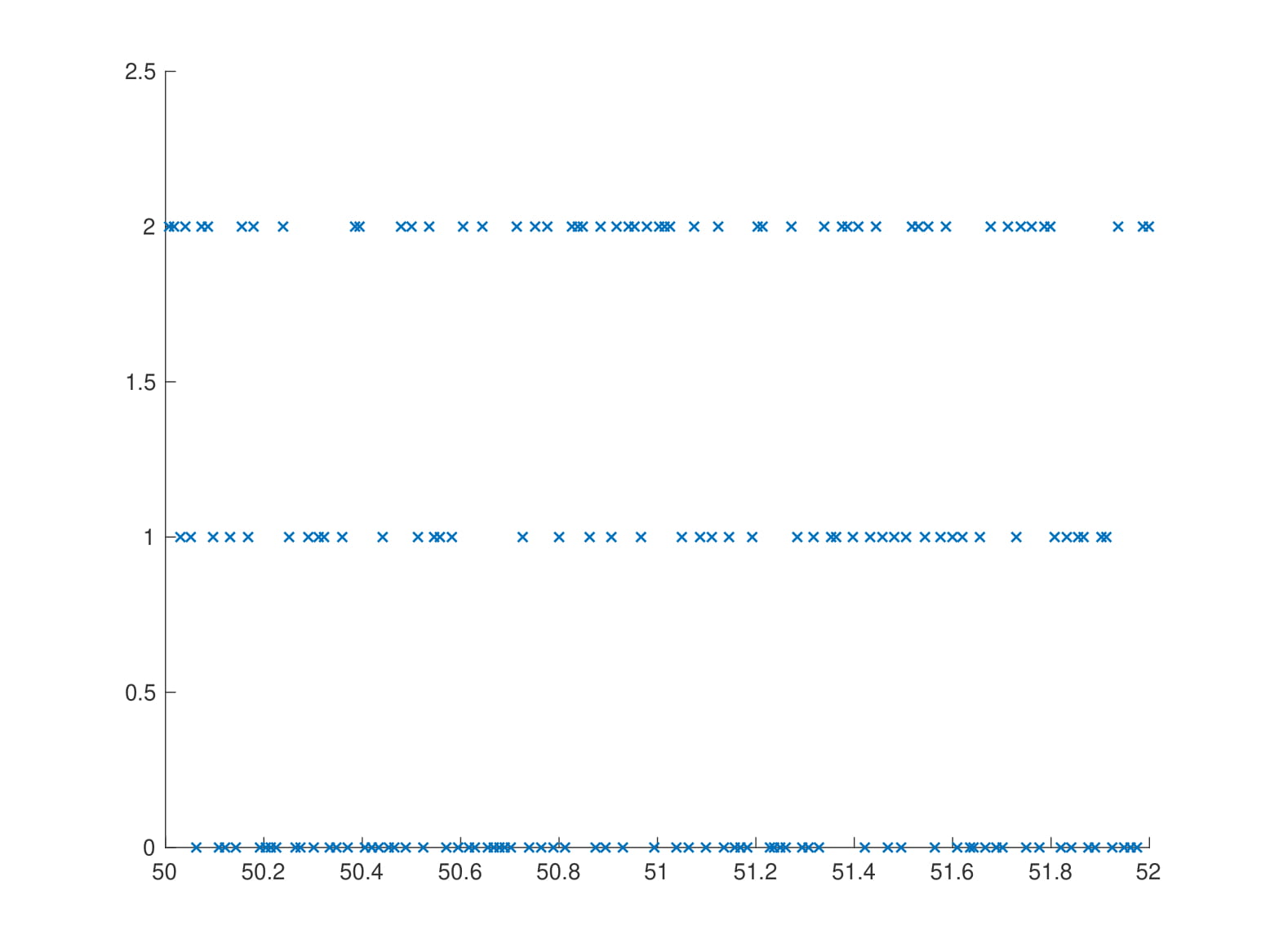,width=0.270\textwidth}}
\caption{Scenario 8}\label{S15}
\end{figure*}

\begin{figure*}[!ht]
\centering%
\subfigure[\footnotesize Output (blue), reference trajectory (red), joystick (green)]
{\epsfig{figure=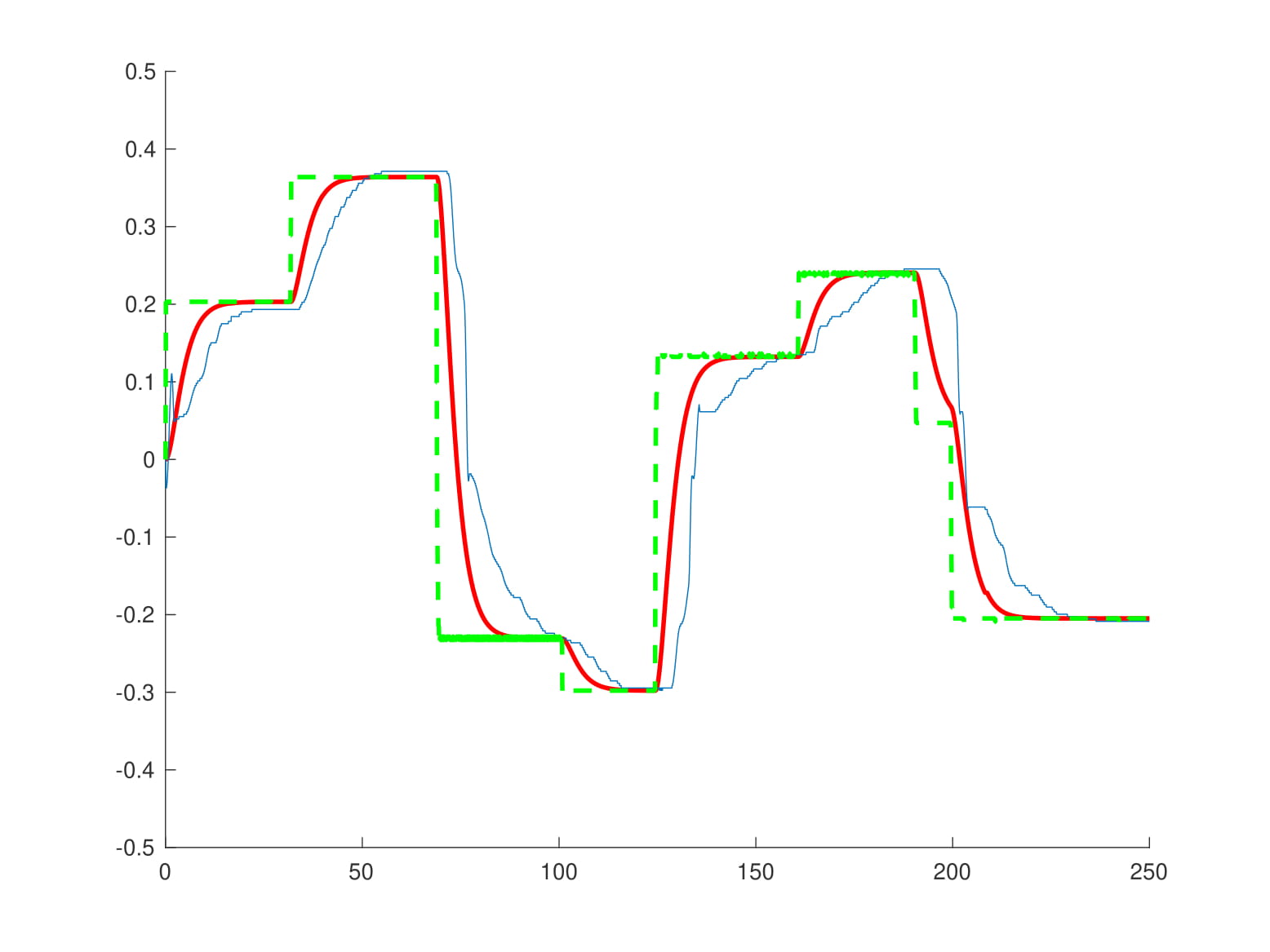,width=0.270\textwidth}}
\subfigure[\footnotesize Supply voltages $v_1$ (blue), $v_2$ (red) ]
{\epsfig{figure=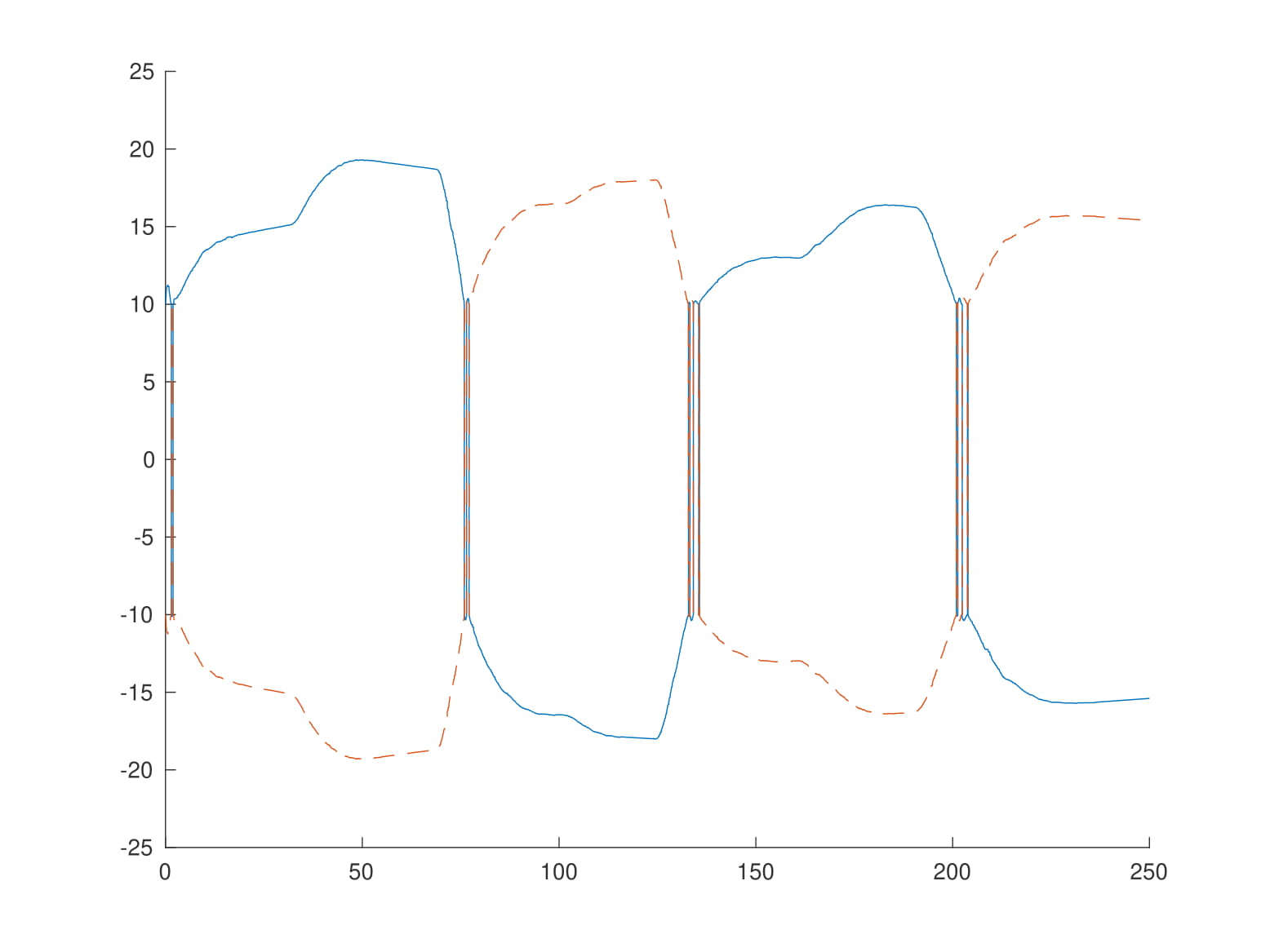,width=0.270\textwidth}}
\subfigure[\footnotesize Zoom on the faults: \{0,1,2\} = \{no fault, fault 1, fault 2\}]
{\epsfig{figure=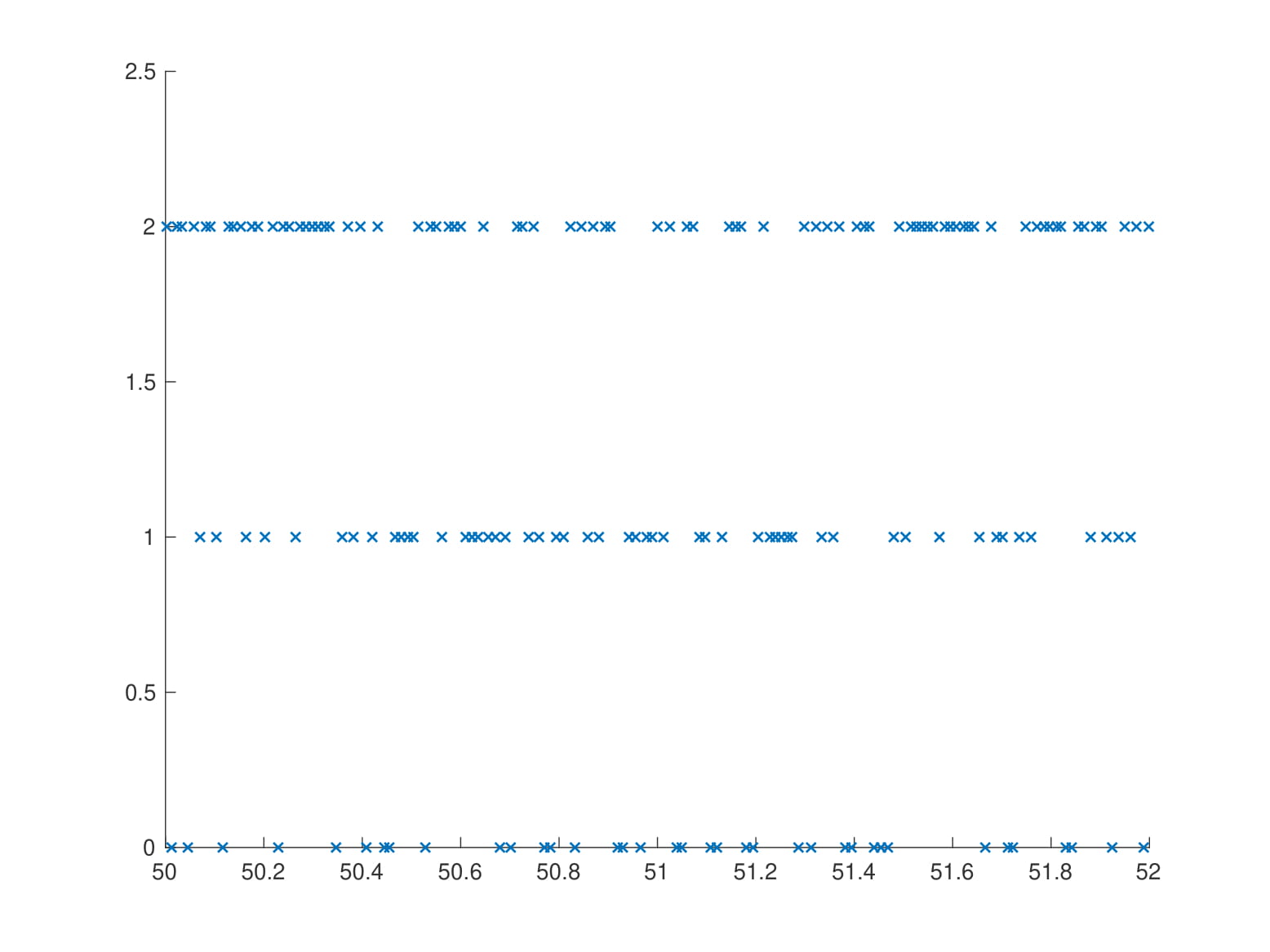,width=0.270\textwidth}}
\caption{Scenario 9}\label{S16}
\end{figure*}

Consider therefore the three following scenarios:
\begin{itemize}
\item {\bf Scenario 7} -- $T=2$s, 2 long transmission cuts with fault 1, and 1 with fault 2; 
\item {\bf Scenario 8} -- $T=2$s, between the supervisor and the server $23.56\%$ of faults 1 and $25.27000\%$ of faults 2;
\item {\bf Scenario 9} -- $T=2$s, between the supervisor and the server $38.79\%$ of faults 1 and $40.50\%$ of fault 2.
\end{itemize}
The results in Figure \ref{S14} are good outside the transmission cuts. Those in Figure \ref{S15} are quite correct in spite of an important packet loss. The scenario 9, where the packet loss is huge, is inducing some lack of efficiency as shown in in Figure \ref{S16}. 

\section{Conclusion}\label{conclusion}
MFC  \cite{csm} might be a most promising tool for control networking. Our study corroborates \cite{sa}: ``Control in the IoT imposes control-theoretic challenges that we are unlikely to encounter in our usual application domains.''  Let us stress therefore that the robustness of MFC with respect to packet loss is today only a purely empirical fact. In the spirit of ``experimental mathematics'' (see, \textit{e.g.}, \cite{arnold}), a theoretical justification needs to be presented.

When the transmission distance becomes large, for instance between France and USA or China, or between the Earth and the Moon,\footnote{Such long distances might be unusual in the IIoT!} latency may perhaps not be neglected anymore. A straightforward extension of our viewpoint yields to a constant delay (compare, \textit{e.g.}, with \cite{ferrari,millnert,millnert1,yas}). In this context the approach on \emph{supply chain management} in \cite{ham} might be useful.

\end{document}